\documentclass[11pt,a4paper]{article}
\usepackage[utf8x]{inputenc}
\usepackage[T1]{fontenc}
\usepackage{mathptmx} 

\usepackage[margin=25truemm]{geometry}
\usepackage[pdftex]{graphicx} 
\usepackage{calc} 
\usepackage{authblk}
\usepackage{bm}
\usepackage{enumitem} 
\usepackage{braket,mathtools,amsmath,amssymb,subcaption}
\usepackage{authblk}
\usepackage[all]{nowidow} 
\usepackage{url}

\usepackage{lipsum} 

\newcommand{\fl}[1]{^{({#1})}}
\newcommand{\fii}[1]{^{I({#1})}}

\usepackage{framed} 
\usepackage[framed]{ntheorem}
\newframedtheorem{frm-thm}{Theorem}

\def\tht{'t Hooft}

\def\hx{\hat{x}}
\def\hy{\hat{y}}
\def\hz{\hat{z}}
\def\tx{\tau^X}

\def\br{\mathbf{r}}

\def\bx{\mathbf{l}_x}
\def\by{\mathbf{l}_y}
\def\bz{\mathbf{l}_z}
\def\bp{\mathbf{p}}

\def\wde{\wedge}

\def\br{\mathbf{r}}

\def\bx{\mathbf{l}_x}
\def\by{\mathbf{l}_y}
\def\bz{\mathbf{l}_z}
\def\bp{\mathbf{p}}

\def\ex{\frac{\mathbf{e}_x}{2}}
\def\ey{\frac{\mathbf{e}_y}{2}}
\def\ez{\frac{\mathbf{e}_z}{2}}

\def\exx{\mathbf{e}_x}
\def\eyy{\mathbf{e}_y}
\def\bc{\mathbf{c}}
\def\ezz{\mathbf{e}_z}

\usepackage{hyperref}
\hypersetup{colorlinks,linkcolor=blue,citecolor=cyan}
\begin{document} 
\title{Modulated symmetries from generalized Lieb-Schultz-Mattis anomalies}

\author{Hiromi Ebisu$^{1,}$$^2$, Bo Han$^3$, Weiguang Cao$^{4,5}$}

\affil{$^1$Yukawa Institute for Theoretical Physics, Kyoto University, Kyoto 606-8502, Japan}
\affil{$^2$Interdisciplinary Theoretical and Mathematical Sciences Program (iTHEMS)~RIKEN, Wako 351-0198, Japan}
\affil{$^3$ Institute for Theoretical Physics, University of Cologne, Physik, Zülpicher Str. 77, 50937 Köln, Germany}    \affil{$^4$~Center for Quantum Mathematics at IMADA, Southern Denmark University, Campusvej 55, 5230 Odense, Denmark}  
\affil{$^5$~Niels Bohr International Academy, Niels Bohr Institute, University of Copenhagen, Blegdamsvej 17DK-2100 Copenhagen, Denmark}
\maketitle
\thispagestyle{empty}

\begin{abstract}
Symmetries rigidly delimit the landscape of quantum matter. Recently uncovered spatially modulated symmetries, whose actions vary with position, enable excitations with restricted mobility, while Lieb–Schultz–Mattis (LSM) type anomalies impose sharp constraints on which lattice phases are realizable. In one-dimensional a spin chain, gauging procedures have linked modulated symmetry to LSM type anomaly, but a general understanding beyond 1D remains incomplete. We show that spatially modulated symmetries and their associated dipole algebras naturally emerge from gauging ordinary symmetries in the presence of generalized LSM type anomalies. We construct explicit lattice models in two and three spatial dimensions and develop complementary field-theoretic descriptions in arbitrary spatial dimensions that connect LSM anomaly inflow to higher-group symmetry structures governing the modulated symmetries. Our results provide a unified, nonperturbative framework that ties together LSM constraints and spatially modulated symmetries across dimensions.
\end{abstract}

\newpage
\pagenumbering{arabic}

\tableofcontents
\section{Introduction}
Symmetry is a unifying organizing principle across physics: it classifies phases of matter, constrains low-energy dynamics, and often enables model-independent predictions. Recent progress has broadened this notion beyond ordinary on-site global symmetries to include higher form symmetries acting on extended objects and categorical symmetries generated by topological defect operators~\cite{doi:10.1073/pnas.0803726105,Frohlich:2006ch,Feiguin2007,gaiotto2015generalized,Aasen_2016,Chang:2018iay,thorngren2019fusion,JiWen2020categorical}.~\footnote{For a complete reference about generalized symmetries, the readers can refer to the following reviews and lecture notes~\cite{McGreevy:2022oyu,Cordova:2022ruw,Shao:2023gho,Bhardwaj:2023kri,Schafer-Nameki:2023jdn}.}
In parallel, spatially modulated symmetries whose symmetry transformations are position-dependent, have emerged as a fertile setting for unconventional phases, constraints and anomalies.

Initially motivated by the fracton topological phases~\cite{chamon,Haah2011,Vijay}, unconventional topological phases of matter admitting excitations with mobility constraints, \textit{(spatially) modulated symmetry} has been developed, giving us diverse research interests. 
While there are several types of the spatially modulated symmetries, in this work, we particularly focus on \textit{dipole symmetry}, which is associated with conservation of dipole moments~\cite{griffin2015scalar,Pretko:2018jbi,PhysRevX.9.031035,Pretko_dis,Gorantla:2022eem}. Key insight of the dipole symmetry is that a mobility constraint is imposed on a single charge due to the conservation of the dipole, leading to interesting physical consequences. The dipole symmetry has played pivotal roles in many aspects of physics. For instance, a new type of Bose-Hubbard model with dipole conserving system reveals rich exotic phases~\cite{LakeHermeleSenthil2201,LakeLeeHanSenthil2210,ZechmannAltmanKnapFeldmeier2210}. Meanwhile, dipole conserving systems show unusual ergodicity breaking  properties, providing a new insight in the context of the eigenvalue thermalization hypothesis~\cite{Sala2022sym,Moudgalya_2022}. 
In addition, there have been growing interests in gauge theory with the modulated symmetries~\cite{Pretko_dis,PhysRevB.106.045145}, including studying theories of anyons with dipole symmetries~\cite{PhysRevB.106.045145,delfino2023anyon,2023foliated}, especially relation to the symmetry enriched topological phases~\cite{PhysRevB.65.165113,PhysRevB.87.104406,PhysRevB.87.155115,Cheng:2015kce,2024multipole,Pace:2025hpb,Tam:2020ifj,Tam:2021efj}, and their anomalies~\cite{dSPT,anomaly_2024,Bulmash2508,Yao:2025iia}.  \par
\begin{table}[h]
\begin{center}
\begin{tabular}{ |c|c|c| c|c|} 
 \hline
 Dim & Global symmetries & Anomaly & Dipole sym by gauging $U_X^{(p)}$&  Dipole sym by gauging $U_Z^{(q)}$\\\hline\hline
 1D & $U_X^{(0)}$, $U^{(0)}_Z$ & $\omega^L$& $0$-form~$\xrightarrow[]{T}$~$0$-form & $0$-form~$\xrightarrow[]{T}$~$0$-form \\ \hline
2D & $U_X^{(0)}$, $U^{(0)}_Z$ & $\omega^{L^2}$& $0$-form~$\xrightarrow[]{T}$~$1$-form & $0$-form~$\xrightarrow[]{T}$~$1$-form \\ \hline
2D & $U_X^{(0),I}$($I=1,2$), $U^{(1)}_Z$& $\omega^L$
&$1$-form~$\xrightarrow[]{T}$~$1$-form & $0$-form~$\xrightarrow[]{T}$~$0$-form \\ \hline
3D & $U_X^{(0),I}$($I=1,2,3$), $U^{(1)}_Z$& $\omega^{L^2}$
&$1$-form~$\xrightarrow[]{T}$~$2$-form & $0$-form~$\xrightarrow[]{T}$~$1$-form \\
 \hline
 \end{tabular} 
    \caption{Summary of our results and previous known fact in the case of $1$D (the first row~\cite{Seifnashri:2023dpa,Pace:2024tgk,Cao:2024qjj}) for comparison. We construct spin models with two types of global symmetries, $U_X^{(p)}$, $U_Z^{(q)}$, comprised of $\mathbb{Z}_N$ Pauli $X$'s and $Z$'s, respectively. The superscript $p$ and $q$ denotes form of the symmetries.  The spin models are anomalous in the sense that two global symmetries, $U_X^{(p)}$, $U_Z^{(q)}$, exhibit unusual commutation relation which depends on $\omega^{L^s}$ with $\omega\vcentcolon=e^{2\pi i/N}$ and $L$ being linear system size. After gauging either~$U_X^{(p)}$ or~$U_Z^{(q)}$, we obtain dipole symmetry, described by a dipole algebra, consisting of $p^\prime$-form and $q^\prime$-form symmetries, the latter of which is generated by acting a translational operator (represented by ``$T$" in the fourth and fifth column ) on the former.  }
    \label{tab:placeholder}
\end{center}
\end{table}
Based on these motivations and backgrounds, in this paper, we try to address the following question: ``how do these modulated symmetries emerge?'' Such a question was partially answered in the case of $1$D spin chain~\cite{Seifnashri:2023dpa,Aksoy:2023hve,Cao:2024qjj}~\footnote{Throughout this paper, the "D" stands for spatial dimension. }: the modulated symmetry emerges from an anomalous spin chain with two global symmetries that exhibit nontrivial commutation relation, depending on the system size. In literature, such an anomaly is referred to as the Lieb-Schultz-Mattis~(LSM) anomaly~\cite{LIEB1961407,1986LMaPh..12...57A,PhysRevLett.84.1535,PhysRevB.69.104431} the name of which comes from the well known theorem that can be used, for instance, to rule out trivial gapped states of matter in systems with a spin-$1/2$ degree of freedom (d.o.f) per unit cell. 
In particular,~$\mathbb Z_N$ dipole symmetry has been obtained 
by gauging one $\mathbb{Z}_N$ global symmetry in a spin chain with the LSM anomaly with respect to $\mathbb{Z}_N\times \mathbb{Z}_N$ global symmetry and translational symmetry. 
However, full exploration of the emergence of modulated symmetries from anomalous quantum spin systems, especially in higher spatial dimensions or involving higher form symmetries, has yet to be investigated. 

In this work, we study various spin models defined in two and three spatial dimensions with two types of global symmetries, exhibiting nontrivial commutation relations which depends on the system size. 
More explicitly, we introduce a~$\mathbb{Z}_N$~spin system with $p$- and~$q$-form  global symmetries, denoted by $U_X\fl p$ and $U_Z\fl q$, defined in $(d+1)$~spacetime dimension,  which is subject to the relation
\begin{eqnarray*}
  U_Z\fl qU_X\fl p= \omega^{L^s} U_X\fl p U_Z\fl q.
\end{eqnarray*}
Here, $L$ represents linear system size and $s=d-p-q$. We show that modulated symmetries are generated by gauging one of the global symmetries~\footnote{Throughout this work, we discuss emergence of the dipole symmetry, which is the simplest example of modulated symmetries. Hence, in the rest of this work, we use the term modulated symmetry and dipole symmetry interchangeably.  }. Furthermore, these modulated symmetries form unconventional dipole algebra -- $p$-form and $q$-form symmetry operators are related with one another via translational operators. 
We emphasize that depending on the system and spatial dimension, $p$ and $q$ are not necessarily the same, which is not discussed in the previous studies.

Our work provides a new insight of emergence of modulated symmetries in a concrete quantum lattice model with generalized LSM type anomaly, making better understanding of these exotic symmetries, especially the ones in spatial dimension more than one. 
We summarize our results in Table.~\ref{tab:placeholder}.
We also give an interpretation of our results by field theoretical analysis, allowing us to understand the emergence of the modulated symmetries in view of (foliated) anomaly inflow terms. The summary of the field theoretical analysis is given in Table.~\ref{222} (Sec.~\ref{5.5}). 
\par
The rest of this work is organized as follows. In Sec.~\ref{sec2}, we review examples of modulated symmetries in 1D and 2D, and how dipole symmetry emerges in 1D. In Sec.~\ref{sec3}, we introduce two anomalous lattice models in 2D to show how modulated symmetries are generated from gauging. In Sec.~\ref{5_1}, we make an interpretation of our results based on the gauge fields associated with dipole symmetries and foliated anomaly inflow terms. Finally, in Sec.~\ref{sec5}, we conclude our work with a few remarks. Technical details, including a thorough analysis of a 3D example, are relegated into appendices.

\section{Review of modulated symmetries}\label{sec2}
In this section, we review modulated symmetries in 1D and 2D lattice models, with a focus on dipole symmetries mixed with lattice translation and ordinary symmetry of the same form. In 1D spin chain, we show that dipole symmetry can emerge after gauging a subgroup of the internal symmetry with a LSM type anomaly. We perform the gauging procedure on the lattice systematically, 
which extends naturally to arbitrary dimensions for generating new modulated symmetries in later sections.
\subsection{Review of dipole symmetry in 1D}\label{1dmd}
\subsubsection{Dipole Ising model and global symmetries}
We start from the simplest spin model that respects modulated symmetry -- 1D dipole Ising model~\cite{Pace:2024tgk,Cao:2024qjj,Ebisu_BO_2025}. The Hamiltonian on a periodic chain with system size $L$  is given by
\begin{eqnarray}
    H_{1D,dipole}=-J\sum_{j=1}^L Z_{j-1}(Z_j^{\dagger})^2 Z_{j+1}-h\sum_{j=1}^L X_j+h.c.,\label{spin}
\end{eqnarray}
where $X_j$ and $Z_j$ are ~$\mathbb{Z}_N$ shift and clock operators at site $j$ with standard relations $Z_jX_j=\omega X_jZ_j$, where $\omega =e^{2\pi i/N}$, $X_j^N=Z_j^N=1$, 
and \textit{h.c.} stands for the Hermitian conjugate.
When $L$ is divisible by $N$,
the model~\eqref{spin} respects the full $\mathbb Z_N\times \mathbb Z_N$  symmetry generated by
\begin{eqnarray}
    Q_0=\prod_{j=1}^{L}X_{j},\quad Q_x=\prod_{j=1}^{L}(X_{j})^j,\label{dipole sym}
\end{eqnarray}
where $Q_0$ generates the ordinary $\mathbb Z_N$ 0-form symmetry and $Q_x$ generates the $\mathbb Z_N$ 0-form dipole symmetry. 
These two global symmetries, including the lattice translational operator $T_x$, form the \textit{dipole algebra}
\begin{eqnarray}
    T_x Q_x T_x^{-1}=Q_0^{\dagger}Q_x,\quad  T_x Q_0 T_x^{-1}=Q_0,\label{0}
\end{eqnarray}
where $T_x$ acts on a local operator $O_j$ by shifting one lattice site as $T_x O_jT^{-1}_x=O_{j+1}$.
Dipole algebra, imposing restricted mobility for single excitations, plays an important role in the context of fracton physics. One can also intuitively construct gauge invariant operators in a gauge theory with modulated symmetries from the dipole algebra~\cite{Ebisu:2023idd,Ebisu:2024eew}. 

\subsubsection{1D dipole symmetry from LSM type anomaly}
\begin{figure}
    \begin{center}
\includegraphics[width=0.4\textwidth]{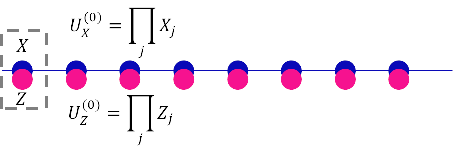}
\end{center}
 \caption{Visual illustration of the LSM anomaly in the 1D chain~\eqref{spin1_d}. On each site, we have a projective representation of $\mathbb{Z}_N\times \mathbb{Z}_N$, i.e. $Z_j X_j = \omega X_j Z_j$.   }\label{1dlsm}
 \end{figure}

In this subsection, we show how the 1D dipole symmetry~\eqref{dipole sym} emerges from gauging a subgroup of internal symmetry with an LSM type anomaly~\cite{Seifnashri:2023dpa,Aksoy:2023hve,Cao:2024qjj}. 
A typical example with an LSM type anomaly in 1D is the XZ model
\begin{equation}
    H_{1D}=-h_x\sum_{j}X^\dagger_jX_{j+1}-h_z\sum_{j}Z^\dagger_jZ_{j+1}+h.c.,\label{spin1_d}
\end{equation}
with $\mathbb Z_N\times \mathbb Z_N$ 0-form symmetry generated by
\begin{eqnarray}
    U_X\fl0\vcentcolon=\prod_{j=1}^L X_j,\quad   U_Z\fl0\vcentcolon=\prod_{j=1}^L Z_j.
\end{eqnarray}
These two symmetry operators exhibit a nontrivial commutation relation
\begin{equation}
    U_Z\fl0U_X\fl0=\omega^LU_X\fl0U_Z\fl0,\label{lsm1}
\end{equation}
with an anomalous phase $\omega^L$ depending on size $L$. 
The relation~\eqref{lsm1} is an indication of the LSM type anomaly -- a mixed anomaly involving  internal symmetry,~$\mathbb{Z}_N\times\mathbb{Z}_N$ in this case, and lattice translation symmetry~\cite{Alavirad:2019iea,Seifnashri:2023dpa,Aksoy:2023hve}. The LSM anomaly is manifested by the projective representation of $\mathbb{Z}_N\times\mathbb{Z}_N$ at each site, as shown in Fig.~\ref{1dlsm}. 
\par
Now we gauge the global symmetry generated by $U_X\fl0$. Generally, gauging is a procedure to promote a global symmetry to a local one~\cite{Shavit_RevModPhys.52.453,PhysRevB.86.115109}. To this end, we introduce extended Hilbert space on each link between adjacent sites, and $\mathbb{Z}_N$ Pauli operators $\widehat{\tau}^X_{j-1/2}, \widehat{\tau}^Z_{j-1/2}$ acting on link $j-1/2$.
We define the Gauss's law as follows:
\begin{eqnarray}
   \widehat{\tau}^Z_{j-1/2} X_j \widehat{\tau}^{Z\dagger}_{j+1/2}=1.\label{gauss_1d}
\end{eqnarray}
Intuitively, the form of the Gauss's law~\eqref{gauss_1d} comes from the fact that one crops the global symmetry $U_X\fl0$ into small segments with inclusion of operators acting on the extended Hilbert space. Indeed, when we multiply $ \widehat{\tau}^Z_{j-1/2} X_j \widehat{\tau}^{Z\dagger}_{j+1/2}$ in the entire space, we reproduce the original global symmetry, viz 
\begin{equation*}
    \prod_{j=1}^L   (\widehat{\tau}^Z_{j-1/2} X_j \widehat{\tau}^{Z \dagger}_{j+1/2})=U_X\fl0.
\end{equation*}

To proceed, we modify the spin coupling term $Z_jZ^{\dagger}_{j+1}$ so that it commutes with the Gauss's law~\eqref{gauss_1d}, corresponding to minimal coupling in the standard gauge theory:
\begin{eqnarray}
    Z^\dagger_jZ_{j+1}\to  Z^\dagger _j\widehat{\tau}^{X\dagger}_{j+1/2}Z_{j+1}.
\end{eqnarray}
After relabeling the operators as
\begin{eqnarray}
\tau^Z_{j+1/2}\vcentcolon=\widehat{\tau}^Z_{j+1/2},\quad \tau^X_{j+1/2}\vcentcolon=Z_j\widehat{\tau}^{X}_{j+1/2}Z^{\dagger}_{j+1},
\end{eqnarray}
the gauged Hamiltonian reads as
\begin{eqnarray}
    \widehat{H}=-h_x\sum_{j}\tau^Z_{j-1/2}(\tau^{Z\dagger}_{j+1/2})^2\tau^Z_{j+3/2}-h_z\sum_j \tau^X_{j+1/2}+h.c..\label{gauge_1d}
\end{eqnarray}
In summary, the gauging procedure induces the following transformation on $\mathbb Z_N$ symmetric operators
\begin{equation}\label{eq:gaugingmap1d}
    Z^\dagger_jZ_{j+1}\to \tau^{X\dagger}_{j+1/2},\quad X_j\to \tau^{Z\dagger}_{j-1/2} \tau^{Z}_{j+1/2}.
\end{equation}
Up to shifting the spin operators by a half-lattice spacing, the gauged model~\eqref{gauge_1d} is identical to~\eqref{spin} and admits the dipole symmetry generated by~\footnote{We get a sequence of theories (labeled by length $L$) after gauging in a theory with the LSM anomaly. At every $L$ as multiple of $N$, the dual theory exhibits full dipole symmetry, while for generic value of $L$ the dipole symmetry may be broken due to the periodic boundary condition. Nevertheless, for every $L$, we can still define dipole symmetry as a bundle symmetry defined on patches covering the whole chain~\cite{Han:2023fas}.}
\begin{eqnarray}
\tilde{Q}_0=\prod_{j=1}^{L}\tau^X_{j+1/2},\quad \tilde{Q}_x= \left[\prod_{j=1}^{L}(\tau^X_{j+1/2})^j \right]^{\alpha}, 
\end{eqnarray}
with $\alpha\vcentcolon=N/\gcd(N,L)$, where gcd stands for the greatest common devisor. $\tilde{Q}_{0}$ is the dual $\mathbb Z_N$ symmetry and $\tilde{Q}_{x}$ arises from $U_{Z}^{(0)}$ by imposing transformation~\eqref{eq:gaugingmap1d} and the periodic boundary condition.
When $L=0~\text{mod}~ N$, we get the full dipole symmetry  identical to~\eqref{dipole sym}. 

\subsection{Examples of modulated symmetries in 2D and dipole algebra}\label{p}
In this subsection, we briefly review modulated symmetries in 2D with a focus on $0$-form and $1$-form dipole symmetries. Here we work on a 2D square lattice with $L_x\times L_y$ sites and assume $L_x=L_y= 0~\text{mod}~N$ to have the full modulated symmetries.

\subsubsection{$0$-form modulated symmetry}
\begin{figure}
    \begin{center}
         \begin{subfigure}[h]{0.49\textwidth}
       \centering
  \includegraphics[width=0.9\textwidth]{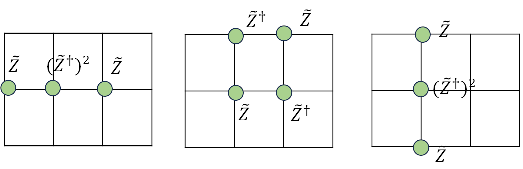}
         \caption{}\label{dualdipole}
             \end{subfigure}
            \begin{subfigure}[h]{0.49\textwidth}
            \centering
  \includegraphics[width=1.0\textwidth]{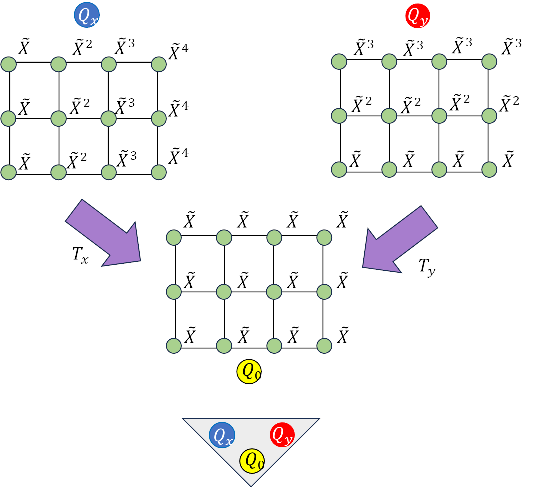}
         \caption{}\label{0formdual}
             \end{subfigure}
 \end{center}
 \caption{(a)~Three types of terms defined in~\eqref{spin2d2} that respecting the $0$-form dipole symmetry~\eqref{algebra}. 
 (b)~$0$-form dipole symmetry, forming dipole algebra~\eqref{algebra} which is schematically portrayed as an inverse of a triangle in the bottom.
 }
 \end{figure}
On the 2D square lattice, we define a $\mathbb{Z}_N$ spin on each site, whose shift and clock operators are given by $X_{\br}, Z_{\br}$. We start by the following Hamiltonian~\cite{Ebisu_BO_2025}
\begin{eqnarray}
    H_{2D}=-J_x\sum_{\br}\mathcal{N}^{Z}_{x,\br}-J_y\sum_{\br}\mathcal{N}^Z_{y,\br}-J_{xy}\sum_{\bp}\mathcal{P}^Z_{\bp}-h\sum_{\br}X_{\br}+h.c.,\label{spin2d}
\end{eqnarray}
with each term defined as
 \begin{equation}
    \mathcal{N}^Z_{x,\br}\vcentcolon=Z_{\br-\exx}(Z^\dagger_{\br})^2Z_{\br+\exx},\,    \mathcal{N}^Z_{y,\br}\vcentcolon=Z_{\br-\eyy}(Z^\dagger_{\br})^2Z_{\br+\eyy},\, 
   \mathcal{P}^Z_{\bp}\vcentcolon=Z^\dagger_{\bp-\ex-\ey}Z_{\bp+\ex-\ey}Z^\dagger_{\bp+\ex+\ey}Z_{\bp-\ex+\ey}\label{spin2d2}.
\end{equation}
where $\br\vcentcolon=(\hat{x},\hat{y}), \hat{x},\hat{y}\in \mathbb{Z}$, and 
$\exx\vcentcolon=(1,0)$, $\exx\vcentcolon=(0,1)$, $\bp\vcentcolon=(\hat{x}+\frac{1}{2},\hat{y}+\frac{1}{2})$.
The spin coupling terms are shown in  Fig.~\ref{dualdipole}.
This Hamiltonian~\eqref{spin2d} respects $\mathbb Z_N\times \mathbb Z_N\times \mathbb Z_N$ 0-form symmetry generated by
\begin{eqnarray}
    Q_{2D:0}=\prod_{\hx=1}^{L_x}\prod_{\hy=1}^{L_y}X_{\br},\quad Q_{2D:x}=\prod_{\hx=1}^{L_x}\prod_{\hy=1}^{L_y}(X_{\br})^{\hx},\quad Q_{2D:y}=\prod_{\hx=1}^{L_x}\prod_{\hy=1}^{L_y}(X_{\br})^{\hy}\label{algebra},
\end{eqnarray}
where $Q_{2D:x}$ and $Q_{2D:y}$ exhibits spacial modulation in $x$- and $y$- directions. They form the following 0-form dipole algebra 
\begin{align}
    T_xQ_{2D:x}T_x^{-1}&=Q_{2D:x}Q_{2D:0}^\dagger,\quad
    T_yQ_{2D:y}T_y^{-1}=Q_{2D:y}Q_{2D:0}^\dagger, \nonumber\\
    T_xQ_{2D:0}T_x^{-1}&= T_yQ_{2D:0}T_y^{-1}=Q_{2D:0},\quad  T_xQ_{2D:y}T_x^{-1}= Q_{2D:y},\quad T_yQ_{2D:x}T_y^{-1}= Q_{2D:x},
    \label{algebra2}
\end{align}
where $T_{I}$ denotes lattice translational operator in the $I$th-direction with the action $T_IO_{\br}T_I^{-1}=O_{\br+e_{I}}$ on local operator $O_{\br}$.
The symmetry operators~\eqref{algebra} realize
the lattice analog of the dipole symmetry in the field theory~\cite{2023foliated}. As shown in \ref{0formdual} there is a hierarchy between the dipole charges $Q_{2D:x},Q_{2D:y}$ and the global one $Q_{2D:0}$: 
the global charge is generated by acting the lattice translational operator on the corresponding dipole charge.

\subsubsection{$1$-form modulated symmetry}
\begin{figure}
    \begin{center}
         \begin{subfigure}[h]{0.49\textwidth}
       \centering
  \includegraphics[width=0.7\textwidth]{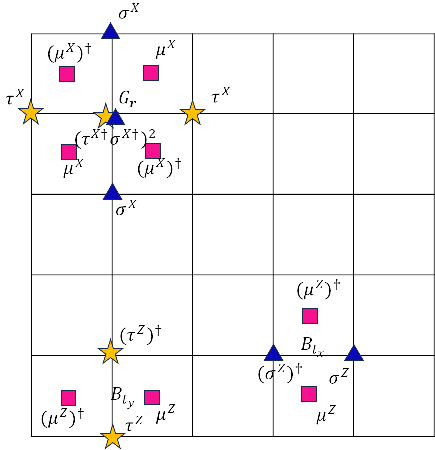}
         \caption{}\label{R1}
             \end{subfigure}
            \begin{subfigure}[h]{0.49\textwidth}
            \centering
  \includegraphics[width=1.2\textwidth]{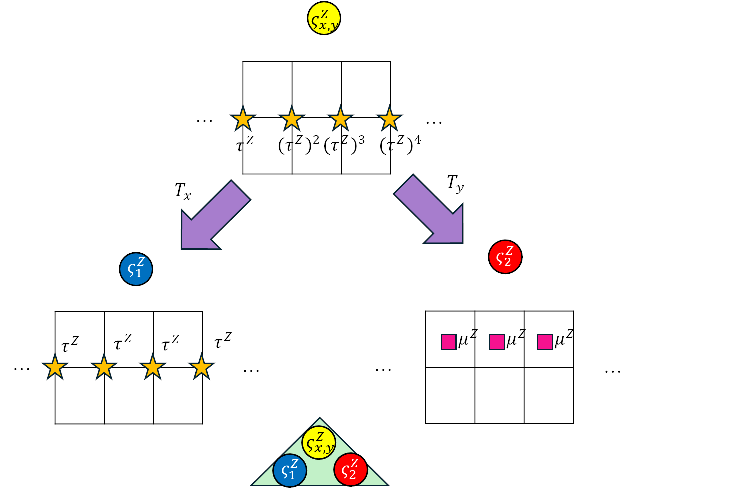}
         \caption{}\label{R2}
             \end{subfigure}
 \end{center}
 \caption{(a)~Three types of terms defined in~\eqref{terms} which constitute the Hamiltonian~\eqref{89}.
 (b)~$1$-form dipole symmetry, forming dipole algebra~\eqref{dual94} which is schematically portrayed as a triangle in the bottom.
 }
 \end{figure}

For $1$-form modulated symmetry, we introduce the rank-2 toric code with dipole symmetry~\cite{PhysRevB.106.045145}. Introducing two sets of $\mathbb{Z}_N$ spins $\tau^{X/Z}_{\br}$, $\sigma^{X/Z}_{\br}$ on each site and another set of $\mathbb{Z}_N$ spins $\mu^{X/Z}_{\bp}$ on each plaquette, the Hamiltonian is described by 
\begin{eqnarray}
\widehat{H}_{DTC}\vcentcolon=-{h}\sum_{\br}G_{\br} - g_{B_x}\sum_{\bx}B_{\bx}-g_{B_y}\sum_{\by}B_{\by}+h.c.\label{89}
\end{eqnarray}
with
\begin{eqnarray}
G_{\br}\vcentcolon&=&{\tx}_{\br-\exx}({\tau}^{X\dagger}_{\br})^2{\tx}_{\br+\exx}\times {\sigma}^X_{\br-\eyy}({\sigma}^{X\dagger}_{\br})^2 {\sigma}^X_{\br+\eyy}\times ({\mu}^{X}_{\bp-\ex+\ey})^\dagger{\mu}^{X}_{\bp-\ex-\ey}({\mu}^X_{\bp+\ex-\ey})^\dagger{\mu}^X_{\bp+\ex+\ey},\nonumber\\
B_{\bx}\vcentcolon&=&\sum_{\bx}\sigma^{Z\dagger}_{\bx-\ex}\sigma^Z_{\bx+\ex}\mu^Z_{\bx-\ey}\mu^{Z\dagger}_{\bx+\ey},\nonumber\\
    B_{\by}\vcentcolon&=&\sum_{\by}\mu^{Z\dagger}_{\by-\ex}\mu^Z_{\by+\ex}\tau^Z_{\by-\ey}\tau^{Z\dagger}_{\by+\ey},\label{terms}
\end{eqnarray}
where
$\bx\vcentcolon=(\hx+\frac{1}{2},\hy)$ and $\by\vcentcolon=(\hx,\hy+\frac{1}{2})$. The terms given in~\eqref{terms} are portrayed in Fig.~\ref{R1}.
\par
Similar to the standard toric code, the model~\eqref{89} is exactly solvable as all terms in the Hamiltonian commute. As such, the ground state is a projected state~$\ket{\omega}$, satisfying, 
\begin{equation}
G_{\br}\ket{\omega}=B_{\bx}\ket{\omega}=B_{\by}\ket{\omega}=\ket{\omega},\quad\forall~\br,\bx,\by.\label{omega}
\end{equation}
On a torus, this Hamiltonian commutes with noncontractible loops of the operators $\tau^Z_{\br}$, $\sigma^Z_{\br}$, and $\mu^Z_{\bp}$ in the $x$-direction as well as the ones in the $y$-direction. There are three types noncontractible loops in the $x$-direction
\begin{eqnarray}
    \xi^Z_{1}\vcentcolon=\prod_{\hx=1}^{L_x}\tau^Z_{(\hx,0)},\quad \xi^Z_2\vcentcolon=\prod_{\hx=1}^{L_x}\mu^Z_{(\hx+\frac{1}{2},\frac{1}{2})},\quad \xi^Z_{x,y}\vcentcolon=\prod_{\hx=1}^{L_x}\left(\tau^{Z}_{(\hx,0)}\right)^{\hx},\label{loops}
\end{eqnarray}
where the last one exhibits spacial modulation.
These loops are depicted in Fig.~\ref{R2}. 
  Note that these operators are topological, i.e., independent operators depend solely on the homology class due to the condition~\eqref{omega}.
A simple calculation, jointly with~\eqref{omega} leads to that
%
\begin{align}
T_x\xi^Z_{x,y}T_x^{-1}&=\xi^{Z\dagger}_1\xi^Z_{x,y},\quad&&  T_y\xi^Z_{x,y}T_y^{-1}=\xi^{Z\dagger}_2\xi^Z_{x,y},\nonumber\\
    T_x\xi^Z_iT_x^{-1}&=\xi^Z_i,\quad&& T_y\xi^Z_iT_y^{-1}=\xi^Z_i~(i=1,2),\label{dual94}
\end{align}
which indicates that 
the loop operators form the 
$1$-form analog of the dipole algebra~\eqref{algebra2}~\cite{Ebisu_BO_2025}. Furthermore, these loops
constitute $1$-form \textit{dual dipole algebra}~\cite{anomaly_2024}, that is, one $1$-form dipole charge followed by two $1$-form global charges, symbolically described by a triangle portrayed in the bottom of Fig.~\ref{R2}. The similar $1$-form operators and dual dipole algebra can be found in the noncontractible loops in the~$y$-direction. 
\par
In the following section, we will demonstrate with concrete lattice models that $0$-form and $1$-form modulated symmetries emerge from gauging an internal ordinary symmetry with the LSM-type anomaly.

\section{Emergence of modulated symmetries from the LSM anomaly in 2D models}\label{sec3}
In this section, we employ the gauging method to generate new modulated symmetries in 2D, as well as to explain the known modulated symmetries shown in the previous section. The main idea is to gauge a subgroup of the internal symmetry with the LSM type anomaly. 
In 2D, we focus on LSM type anomalies \textit{generalized} in two directions, comparing to the 1D case. 
In Sec.~\ref{00}, we start from two $0$-form global symmetries whose symmetry operators satisfy an anomalous commutation relation. However, the anomalous phase in this case depends on the \textit{area} of the system. This signals a generalized version of the LSM type anomaly where the anomalous phase usually depends on the {\it linear} size of the system. After gauging, we obtain new modulated symmetry with dipole symmetry mixing with ordinary symmetry of a \textit{different} form.
In Sec.~\ref{32}, we 
study LSM type anomaly involving \textit{different} forms of internal symmetries. We illustrate this by one $0$-form and one $1$-form symmetries where the commutation relation between symmetry operators depends on the length of the system. After gauging either the 0-form or the 1-form symmetry, we obtain various modulated symmetries explored in Sec.~\ref{p}.

\subsection{Two $0$-form symmetries}\label{00}
We consider the following Hamiltonian on a square lattice with system size $L_x\times L_y$ and periodic boundary condition:
\begin{equation}
    H_{2D,0}=-\sum_{\br}(X_\br^\dagger X_{\br+\exx}+X_\br^\dagger X_{\br+\eyy})-\sum_{\br}(Z_\br^\dagger Z_{\br+\exx}+Z_\br^\dagger Z_{\br+\eyy})+h.c. .\label{spin001}
\end{equation}
This model has  $\mathbb Z_N\times \mathbb Z_N$ $0$-form symmetry generated by
\begin{eqnarray}
    U^{(0)}_X=\prod_{\hx=1}^{L_x}\prod_{\hy=1}^{L_y}X_{\br},\quad U^{(0)}_Z=\prod_{\hx=1}^{L_x}\prod_{\hy=1}^{L_y}Z_{\br},\label{global2d}
\end{eqnarray}
with the anomalous commutation relation
\begin{eqnarray}
    U^{(0)}_{Z}U^{(0)}_{X}=\omega^{L_xL_y}U^{(0)}_{Z}U^{(0)}_{X}.
\end{eqnarray}
Closely parallel to the 1D example in~\eqref{lsm1}, this anomalous phase $\omega^{L_xL_y}$, depending on the area of the system, signals a generalized LSM anomaly
involving internal $0$-form global symmetry $\mathbb{Z}_N\times\mathbb{Z}_N$ and lattice translation symmetry in the $x$- and $y$-direction: at each node of the lattice, we have a projective representation of $\mathbb{Z}_N\times\mathbb{Z}_N$. 
\par
\begin{figure}
    
    \hspace{-10mm}
         \begin{subfigure}[h]{0.49\textwidth}
       \centering
  \includegraphics[width=0.7\textwidth]{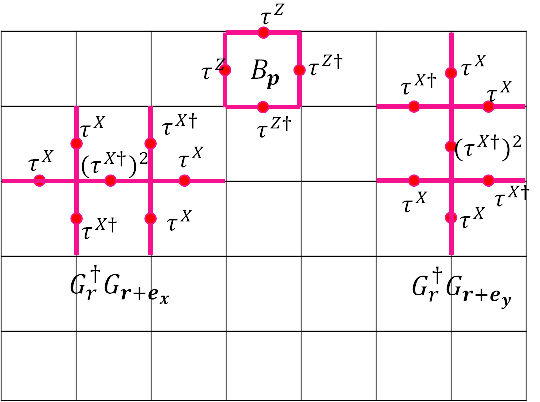}
         \caption{}\label{01}
             \end{subfigure}
            \begin{subfigure}[h]{0.49\textwidth}
            \centering
  \includegraphics[width=0.7\textwidth]{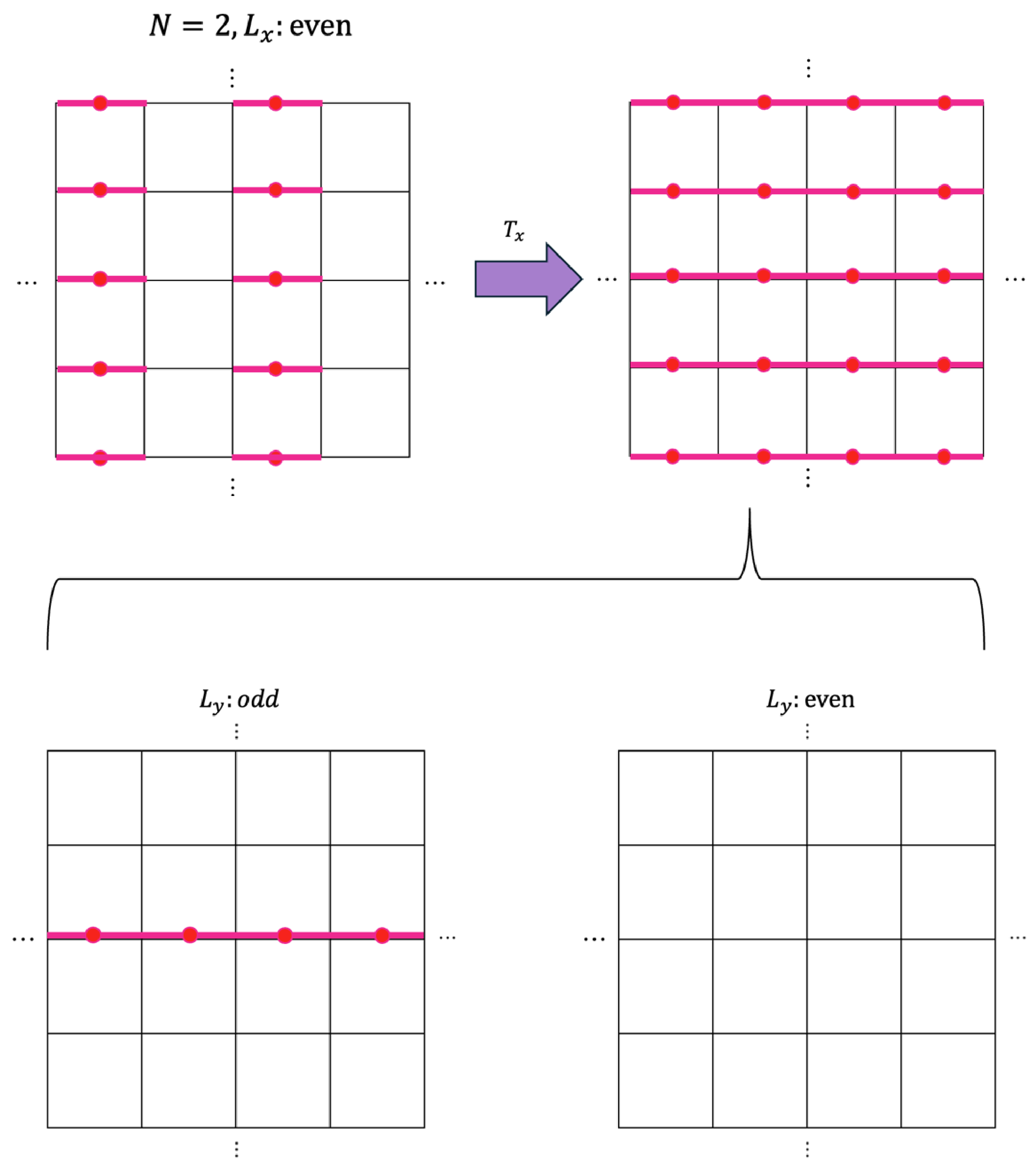}
         \caption{}\label{02}
             \end{subfigure}
 \caption{(a)~The first three terms in~\eqref{2toric}. (b) Example of the dipole algebra~\eqref{dipole01} in the case of $N=2$ and $L_x$ even. Note that as opposed to previous studies which discuss dipole algebra involving the same form, we have unusual dipole algebra  which involves $0$-form and $1$-form symmetry. Namely, acting a translational operator on the $0$-form symmetry yields stack of $1$-form symmetries. In the present case, the stack of $1$-form symmetries can be deformed into identity of one $1$-form symmetry, depending on whether $L_y$ is even or odd via flatness condition of the gauge field. 
 }
 \end{figure}

Now we gauge one of the global symmetries generated by $U^{(0)}_X$~\cite{Shavit_RevModPhys.52.453,PhysRevB.86.115109}. To this end, we accommodate extended Hilbert space on each link of the lattice with a new set of Pauli operators  $\widehat{\tau}^{X/Z}$, 
and impose the following Gauss's law
\begin{eqnarray}
    \widehat{\tau}^{X\dagger}_{\br-\ex}  \widehat{\tau}^{X\dagger}_{\br-\ey}X_{\br}  \widehat{\tau}^{X}_{\br+\ex}\widehat{\tau}^{X}_{\br+\ey}=1.\label{gauss 2d}
\end{eqnarray}
Similar to the 1D case in~\eqref{gauss_1d}, this Gauss's law 
comes from cropping the global symmetry operator $U^{(0)}_X$ into small segments, which can be understood as
\begin{eqnarray*}
    \prod_{\br}(  \widehat{\tau}^{X\dagger}_{\br-\ex}  \widehat{\tau}^{X\dagger}_{\br-\ey}X_{\br}  \widehat{\tau}^{X}_{\br+\ex}\widehat{\tau}^{X}_{\br+\ey})=U^{(0)}_X.
\end{eqnarray*}
To proceed, we modify the spin coupling terms, $Z_\br^{\dagger}Z_{\br+\exx}$ and~$Z_\br^{\dagger}Z_{\br+\eyy}$ 
\begin{eqnarray}
    Z_\br^{\dagger}Z_{\br+\exx}\to   Z_\br^{\dagger}\widehat{\tau}^{Z \dagger}_{\br+\ex}Z_{\br+\exx},\quad  Z_\br^{\dagger}Z_{\br+\eyy}\to   Z_\br^{\dagger}\widehat{\tau}^{Z \dagger}_{\br+\ey}Z_{\br+\eyy},
\end{eqnarray} 
such that they commute with the Gauss's law term.
Furthermore, we add the following flux operator of the gauge field
\begin{eqnarray}
    -g\sum_{\bp}\widehat{\tau}^Z_{\bp-\ex}\widehat{\tau}^{Z\dagger}_{\bp+\ex}\widehat{\tau}^Z_{\bp+\ey}\widehat{\tau}^{Z\dagger}_{\bp-\ey}+h.c.,
\end{eqnarray}
to the Hamiltonian to ensure that the gauged theory is dynamically trivial.
Rewriting operators as
\begin{eqnarray}
    &\tau^{X}_{\br+\ex}\vcentcolon= \widehat{\tau}^{X}_{\br+\ex},\quad  \tau^{X}_{\br+\ey}\vcentcolon= \widehat{\tau}^{X}_{\br+\ey}\nonumber\\
&\tau^Z_{\br+\ex}\vcentcolon=Z_\br^{\dagger}\widehat{\tau}^{Z \dagger}_{\br+\ex}Z_{\br+\exx} \quad \tau^Z_{\br+\ey}\vcentcolon=Z_\br^{\dagger}\widehat{\tau}^{Z \dagger}_{\br+\ey}Z_{\br+\eyy},    \end{eqnarray}
we obtain the following gauged Hamiltonian:
\begin{eqnarray}
    \widehat{H}_{2D,0}=-\sum_{\br}(G^\dagger_{\br}\times G_{\br+\exx}+G^\dagger_{\br}\times G_{\br+\eyy})-g\sum_{\bp} B_{\bp}-\sum_{\bx}\tau^Z_{\bx}-\sum_{\by}\tau^Z_{\by}+
    h.c.,\label{2toric}
\end{eqnarray}
with
\begin{eqnarray}
    G_{\br}\vcentcolon=\tau^{X\dagger}_{\br-\ex}\tau^X_{\br+\ex}\tau^{X\dagger}_{\br-\ey}\tau^{X}_{\br+\ey},\quad B_{\bp}\vcentcolon={\tau}^Z_{\bp-\ex}{\tau}^{Z\dagger}_{\bp+\ex}{\tau}^Z_{\bp+\ey}{\tau}^{Z\dagger}_{\bp-\ey}.
\end{eqnarray}
The first three terms in gauged model~\eqref{2toric}, also shown in Fig.~\ref{01}, resemble the ones constituting the~$\mathbb{Z}_N$
toric code~\cite{KITAEV20032} with a crucial difference that 
we have product of the ``star operators''~$G_{\br}$ in the two consecutive sites.
\par
Now we turn to identifying symmetry of the model~\eqref{2toric} on torus geometry, assuming $g\to\infty$ so that the model does not admit any magnetic flux, i.e., the ground state $\ket{\Omega}$ is a projected state, satisfying
\begin{eqnarray}
    B_\bp\ket{\Omega}=\ket{\Omega},\quad\forall\bp.\label{flux}
\end{eqnarray}
The gauged model~\eqref{2toric} respects a new 0-form modeulated symmetry generated by
\begin{eqnarray} 
Q_{x}^{(0)}\vcentcolon=\prod_{\hy=1}^{L_y}\prod_{\hx=1}^{L_x}\left( [\tau^Z_{\bx}]^{\hx}\right)^{\alpha_{x}},\quad Q_{y}^{(0)}\vcentcolon=\prod_{\hy=1}^{L_y}\prod_{\hx=1}^{L_x}\left( [\tau^Z_{\by}]^{\hy}\right)^{\alpha_{y}},\label{0form}
\end{eqnarray}
where $\alpha_I=N/\gcd(N,L_I),~I=x,y$. 
This modulated symmetry depends on the system size in a subtle way. Note that depending on the $N$ and values of $L_x,L_y$, these two charges are not independent: If $\gcd(N,L_I)>1$, further,  if there exit integers $\{c_I:1\leq c_I\leq \gcd(N,L_I)-1$\} such that $c_x\alpha_x+c_y\alpha_y=0~\text{mod}~N$, then 
    by~\eqref{flux}, it follows that 
    the two charges are subject to
$\left[Q_{x}^{(0)}\right]^{c_x}\times \left[Q_{y}^{(0)}\right]^{c_y} =I$, where $I$ on the right hand side represents identity operator.
\par

Besides the modulated $0$-form symmetries that cover entire horizontal or vertical link in the bulk, the model admits 
ordinary $\mathbb Z_N$ $1$-form symmetries, corresponding to the noncontractible Wilson loops along the $x$- and $y$-direction of the torus, that is,
\begin{eqnarray}
Q_{0}^{(1),x}\vcentcolon=\prod_{\hx=1}^{L_x}\tau^Z_{(\hx+\frac{1}{2},1)},\quad Q_{0}^{(1),y}\vcentcolon=\prod_{\hy=1}^{L_y}\tau^Z_{(1,\hy+\frac{1}{2})}.\label{1form}
\end{eqnarray}
These loops are topological, viz, independent operators depend solely on the homology class due to the flux-less condition~\eqref{flux}.
Moreover, the symmetry operators of these 0-form symmetry and the  1-form symmetry \eqref{1form} generate a dipole algebra, that is, 
\begin{eqnarray}
    T_xQ_x^{(0)}T_x^{-1}=Q_x^{(0)}\left(Q_{0}^{(1),x\dagger}\right)^{\alpha_xL_y},\quad T_yQ_x^{(0)}T_y^{-1}=Q_x^{(0)}\nonumber\\
T_xQ_y^{(0)}T_x^{-1}=Q_y^{(0)},\quad T_yQ_y^{(0)}T_y^{-1}=Q_y^{(0)}\left(Q_{0}^{(1),y\dagger}\right)^{\alpha_yL_x}.\label{dipole01}
\end{eqnarray}
The first and last relation in~\eqref{dipole01} indicates that acting a translational operator on a $0$-form symmetry gives rise to \textit{stack} of $1$-form symmetries in the $x$- or $y$-direction. Since the 1-form symmetry is topological, the stacking is manifest in the exponent $L_y,L_x$. 
The subtlety of this modulated symmetry is also reflected in the dipole algebra. At the extreme case, we have the full modulated feature of $Q_x^{(0)}$ if $L_x=0~\text{mod}~N$ and $L_y=1~\text{mod}~N$. But then the modulation for $Q_y^{(0)}$ lost entirely. If we require $L_x=L_y=0~\text{mod}~N$, all modulated features will lost.
We demonstrate this behavior through the first relation of~\eqref{dipole01} in Fig.~\ref{02} in the case of $N=2$, $L_x$ being even. 

Therefore, using the gauging method, we construct a new type of dipole algebra mixing dipole symmetry with ordinary symmetry of \textit{different} form. The known examples in the literature only cover the dipole algebra where $p$-form symmetry and another $p$-form one are related via lattice translational operators, while in our new example~\eqref{dipole01}, different forms of symmetries are associated with one another by lattice translational operators. 

\subsection{Two $0$-form and one $1$-form symmetries}\label{32}
In this subsection, we demonstrate that our gauging method is able to explain the known modulated symmetries in 2D. We start from two 0-form and one 1-form symmetries with an LSM type anomaly implied by the linear system size dependence in the anomalous commutation relations. After gauging, we recover modulated symmetries discussed in Sec.\ref{p}.

\subsubsection{Model}
\begin{figure}
    \begin{center}
         \begin{subfigure}[h]{0.49\textwidth}
       \centering
  \includegraphics[width=0.5\textwidth]{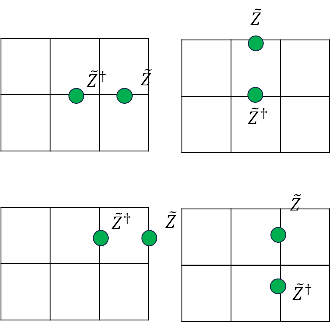}
         \caption{}\label{spinterm}
             \end{subfigure}
            \begin{subfigure}[h]{0.49\textwidth}
            \centering
  \includegraphics[width=0.7\textwidth]{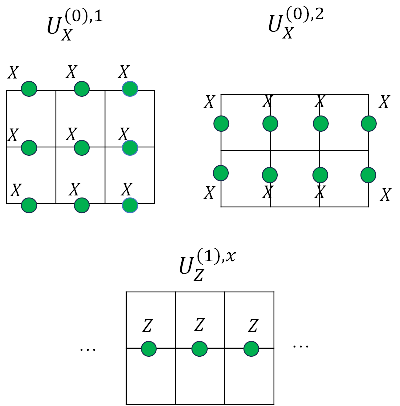}
         \caption{}\label{globalsym}
             \end{subfigure}
 \end{center}
      \begin{subfigure}[h]{0.49\textwidth}
            \centering
  \includegraphics[width=0.6\textwidth]{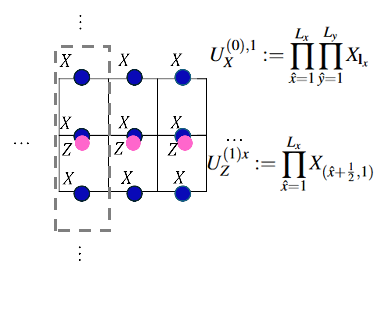}
         \caption{}\label{gauss00}
             \end{subfigure}
             \begin{subfigure}[h]{0.49\textwidth}
            \centering
  \includegraphics[width=0.6\textwidth]{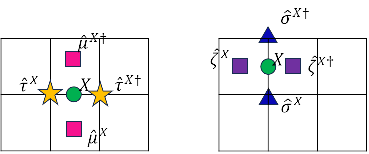}
         \caption{}\label{gauss0}
             \end{subfigure}
          
                \begin{subfigure}[h]{0.49\textwidth}
            \centering
  \includegraphics[width=0.6\textwidth]{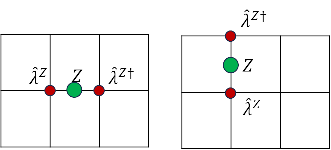}
         \caption{}\label{gauss11}
             \end{subfigure}
 \caption{(a)~Spin coupling terms in the first line of~\eqref{01spin}.
 (b)~(Top)~Two $0$-form symmetries in~\eqref{20form}. (Bottom)~Example of the $1$-form symmetry, corresponding to $U_Z^{(1),x}$ in~\eqref{1_form}. 
 (c) Visual illustration of the LSM anomaly in our model: in each slab (the area inside the gray dashed line, we have two operators, $  Z_{(\hx+\frac{1}{2},1)}$, $\left(\prod_{\hy=1}^{L_y}X_{\bx}\right)$ which do not commute, signaling anomaly involving $0$-form and $1$-form symmetries and translational one in $x$-direction, in analogy to the 1D case (Fig.~\ref{1dlsm}).
 (d) Gauss's law for gauging two 0-form symmetries~\eqref{gauss3}. (e) Gauss's law for gauging $1$-form symmetries~\eqref{gauss4}.
 }
 \end{figure}
We consider $\mathbb{Z}_N$ spin on each link of a 2D square lattice with system size $L_x\times L_y$ and periodic boundary condition, and introduce the Hamiltonian as
 \begin{eqnarray}
     H_{2D,1}=&-&J_x\sum_{\bx}(Z_{\bx}Z^\dagger_{\bx+\exx}+Z_{\bx}Z^\dagger_{\bx+\eyy})-J_y\sum_{\by}(Z_{\by}Z^\dagger_{\by+\eyy}+Z_{\by}Z^\dagger_{\by+\eyy})\nonumber\\
     &-&J_G\sum_{\br}X^{\dagger}_{\br-\ex}X_{\br+\ex}X^{\dagger}_{\br-\ey}X_{\br+\ey}-J_B\sum_\bp Z_{\bp-\ex}Z^{\dagger}_{\bp+\ex}Z_{\bp+\ey}Z^{\dagger}_{\bp-\ey}+h.c.,\label{01spin}
 \end{eqnarray} 
where the first line  describes spin coupling terms shown in Fig.~\ref{spinterm}, and the second line describes the $\mathbb{Z}_N$ toric code. We assume $J_B\to\infty$ so that we focus on the projected states without gauge fluxes. \par
Now we turn to identifying global symmetries of the model. There are two $\mathbb Z_N$ $0$-form global  symmetries generated by 
 \begin{eqnarray}
U^{(0),1}_X\vcentcolon=\prod_{\hx=1}^{L_x}\prod_{\hy=1}^{L_y}X_{\bx},\quad U^{(0),2}_X\vcentcolon=\prod_{\hx=1}^{L_x}\prod_{\hy=1}^{L_y}X_{\by}\label{20form},
\end{eqnarray}
and one $\mathbb Z_N$ $1$-form symmetry generated by noncontractible loop operators along $x$- or $y$-direction
 \begin{eqnarray} U^{(1)x}_Z\vcentcolon=\prod_{\hx=1}^{L_x}Z_{(\hx+\frac{1}{2},1)},\quad U^{(1)y}_Z\vcentcolon=\prod_{\hy=1}^{L_y}Z_{(1,\hy+\frac{1}{2})}.\label{1_form}
 \end{eqnarray}
Note that these loops are topological
due to the flux-less condition by taking $J_B\to\infty$. In Fig.~\ref{globalsym}, we show examples of these symmetry operators. The $0$-form~\eqref{20form} and $1$-form symmetry operators~\eqref{1_form} satisfy the following anomalous commutation relations:
 \begin{eqnarray}
     U^{(1)x}_ZU^{(0),1}_X=\omega^{L_x}U^{(0),1}_X  U^{(1)x}_Z,\quad U^{(1)y}_ZU^{(0),2}_X=\omega^{L_y}U^{(0),2}_X  U^{(1)y}_Z.
 \end{eqnarray}
Symmetry operators with different forms exhibit unusual commutations relation with anomalous phases $(\omega^{L_x},\omega^{L_y})$ depending on system size $(L_x,L_y)$, as opposed to previous cases, where nontrivial commutation relation involves only $0$-form symmetries. This implies the LSM type anomaly: as shown in Fig~\ref{gauss00}, in each slab, $\Omega_{\hx}\in\{(\hx+\frac{1}{2},\hy),~1\leq\hy\leq L_y\}~(1\leq \hx\leq L_x)$, we have nontrivial commutation relations between two operators as
\begin{eqnarray*}
    Z_{(\hx+\frac{1}{2},1)}\left(\prod_{\hy=1}^{L_y}X_{\bx}\right)=\omega Z_{(\hx+\frac{1}{2},1)}\left(\prod_{\hy=1}^{L_y}X_{\bx}\right)\quad (1\leq \hx\leq L_x),
\end{eqnarray*}
signaling the LSM anomaly involving $0$-form and $1$-form and translational symmetry in the $x$-direction. The indication of LSM anomaly in the $y$-direction can be analogously discussed.
As compared between Fig.~\ref{1dlsm} and Fig~\ref{gauss00}, this is generalization of~\eqref{lsm1} where there is a projective representation of $\mathbb{Z}_N\times \mathbb{Z}_N$ at each site of the chain.
In what follows, we gauge either the $0$-form or the $1$-form global symmetries, which leads to different kinds of modulated symmetries in Sec.~\ref{p}.

\subsubsection{Gauging $0$-form symmetry} 
First, we gauge two $\mathbb Z_N$ $0$-form symmetries generated by ~\eqref{20form} in the model~\eqref{01spin}. 
We introduce two copies of 
extended Hilbert spaces on each site and another two on each plaquette of the lattice, with two sets of $\mathbb{Z}_N$ Pauli operators on each site and plaquette as $\widehat{\sigma}^X_{\br}$, $\widehat{\tau}^X_{\br}$ and $\widehat{\mu}^X_{\bp}$, $\widehat{\xi}^X_{\bp}$. Then we introduce the Gauss's law
\begin{eqnarray}
    \widehat{\mu}^X_{\bp-\eyy}\widehat{\tau}^X_{\br}  X_{\bx}\widehat{\tau}^{X\dagger}_{\br+\exx}\widehat{\mu}^{X\dagger}_\bp=1,\quad   \widehat{\xi}^X_{\bp-\exx}\widehat{\sigma}^X_{\br}  X_{\by}\widehat{\sigma}^{X\dagger}_{\br+\eyy}\widehat{\xi}^{X\dagger}_\bp=1\label{gauss3}
\end{eqnarray}
See also in Fig.~\ref{gauss0}.
The Gauss's law terms come from cropping the $0$-form global symmetry operators~\eqref{20form} into local segments. Indeed, by multiplying the Gauss's law terms on every site, we obtain the original 0-form symmetry operators 
\begin{eqnarray*}
    \prod_{\hx=1}^{L_x}  \prod_{\hy=1}^{L_y}  \left(\widehat{\mu}^X_{\bp-\eyy}\widehat{\tau}^X_{\br}  X_{\bx}\widehat{\tau}^{X\dagger}_{\br+\exx}\widehat{\mu}^{X\dagger}_\bp\right)=U_X^{(0),1},\quad \prod _{\hx=1}^{L_x}\prod_{\hy=1}^{L_y}\quad  \left( \widehat{\xi}^X_{\bp-\exx}\widehat{\sigma}^X_{\br}  X_{\by}\widehat{\sigma}^{X\dagger}_{\br+\eyy}\widehat{\xi}^{X\dagger}_\bp\right)=U_x^{(0),2}.
\end{eqnarray*}
We then modify the spin terms in the first line of~\eqref{01spin}
\begin{eqnarray}
Z_{\bx}Z^\dagger_{\bx+\exx}\to Z_{\bx}\widehat{\tau}^{Z\dagger}_{\br}Z^\dagger_{\bx+\exx},\quad Z_{\bx}Z^\dagger_{\bx+\eyy}\to Z_{\bx}\widehat{\xi}^{Z\dagger}_\bp Z^\dagger_{\bx+\eyy}\nonumber\\
Z_{\by}Z^\dagger_{\by+\eyy}\to
Z_{\by}\widehat{\sigma}^{Z\dagger}_\br Z^\dagger_{\by+\eyy},\quad Z_{\by}Z^\dagger_{\by+\eyy}\to Z_{\by}\widehat{\mu}^{Z\dagger}_\bp Z^\dagger_{\by+\eyy}.\label{modify1}
\end{eqnarray} 
so that they commute with Gauss's laws~\eqref{gauss3}.
To proceed, we rewrite the operators as 
\begin{eqnarray}
    &\sigma^X_\br \vcentcolon=\widehat{\sigma}^X_\br,\quad \sigma^Z_\br \vcentcolon=Z_{\by}^\dagger\widehat{\sigma}^{Z}_\br Z_{\by+\eyy},\quad \tau^X_\br\vcentcolon=\widehat{\tau}^X_\br,\quad \tau^Z_\br\vcentcolon=Z_{\bx}^\dagger\widehat{\tau}^{Z}_{\br}Z_{\bx+\exx}\nonumber\\
&\xi_\bp^X\vcentcolon=\widehat{\xi}_\bp^X,\quad \xi^Z_\bp\vcentcolon=Z_{\bx}^\dagger\widehat{\xi}^{Z}_\bp Z_{\bx+\eyy},\quad \mu^X_\bp\vcentcolon=\widehat{\mu}^X_\bp,\quad \mu^Z_\bp\vcentcolon=Z_{\by}^\dagger\widehat{\mu}^{Z}_\bp Z_{\by+\eyy}.\label{modify2}
\end{eqnarray}
We also add the following flux operators 
\begin{eqnarray}
    -J_1\sum _{\bx}\sigma^Z_{\bx+\ex}
\sigma^{Z\dagger}_{\bx-\ex}\mu^{Z\dagger}_{_{\bx+\ey}}\mu^{Z}_{_{\bx-\ey}}-J_2\sum_{\by}\tau^Z_{\by+\ey}\tau^{Z\dagger}_{\by-\ey}\xi^{Z\dagger}_{\by+\ex}\xi^{Z}_{\by-\ex}+h.c.,\label{flux2}
\end{eqnarray}
to the Hamiltonian to make the gauged theory dynamically trivial.
Further, by using Gauss's laws~\eqref{gauss3}, the ``star operator'' of the toric code [third term in~\eqref{01spin}] becomes
\begin{eqnarray}
     X^{\dagger}_{\br-\ex}X_{\br+\ex}X^{\dagger}_{\br-\ey}X_{\br+\ey}&=&   {\tx}_{\br-\exx}({\tau}^{X\dagger}_{\br})^2{\tx}_{\br+\exx}\times {\sigma}^X_{\br-\eyy}({\sigma}^{X\dagger}_{\br})^2 {\sigma}^X_{\br+\eyy}\nonumber\\
     &\times& \left({\mu}^{X}_{\bp-\ex+\ey}{\xi}^{X}_{\bp-\ex+\ey}\right)^\dagger{\mu}^{X}_{\bp-\ex-\ey}{\xi}^{X}_{\bp-\ex-\ey}\nonumber\\
     &\times&\left({\mu}^X_{\bp+\ex-\ey}{\xi}^X_{\bp+\ex-\ey}\right)^\dagger{\mu}^X_{\bp+\ex+\ey}{\xi}^X_{\bp+\ex+\ey}.
\end{eqnarray}
Also, in the original model~\eqref{01spin}, we impose the fluxless condition on the theory, i.e., we focus on the projected state satisfying 
\begin{eqnarray}
    Z_{\bp-\ex}Z^{\dagger}_{\bp+\ex}Z_{\bp+\ey}Z^{\dagger}_{\bp-\ey}=1,\quad \forall\bp.
\end{eqnarray}
After the procedure~\eqref{modify1}~\eqref{modify2}, this condition becomes
\begin{eqnarray}
    \mu^Z_{\bp}\xi^Z_{\bp}=1,\quad \forall \bp.
\end{eqnarray}
Applying this condition to the Hamiltonian to replace all the $\xi_{\bp}$ with $\eta_{\bp}$ and  rewriting the operator~$\mu^{X}_{\bp}\xi^{X}_{\bp}\to \mu^{X}_{\bp}$, we arrive at the gauged Hamiltonian
\begin{eqnarray}
    \widehat{H}_{2D,1}=-J_x\sum (\sigma^Z_{\br}+\mu^Z_{\bp})-J_y\sum (\tau^Z_{\br}+\mu^Z_{\bp})-J_G\sum_{\br}G_{\br}-J_1\sum_{\bx}B_{\bx}-J_2\sum_{\by}B_{\by}+h.c.,
\end{eqnarray}
with
\begin{eqnarray}
   G_{\br}&=&{\tx}_{\br-\exx}({\tau}^{X\dagger}_{\br})^2{\tx}_{\br+\exx}\times {\sigma}^X_{\br-\eyy}({\sigma}^{X\dagger}_{\br})^2 {\sigma}^X_{\br+\eyy}\times ({\mu}^{X}_{\bp-\ex+\ey})^\dagger{\mu}^{X}_{\bp-\ex-\ey}({\mu}^X_{\bp+\ex-\ey})^\dagger{\mu}^X_{\bp+\ex+\ey},\nonumber\\
B_{\bx}&=&\sum_{\bx}\sigma^{Z\dagger}_{\bx-\ex}\sigma^Z_{\bx+\ex}\mu^Z_{\bx-\ey}\mu^{Z\dagger}_{\bx+\ey},\nonumber\\
   B_{\by}&=&\sum_{\by}\mu^{Z\dagger}_{\by-\ex}\mu^Z_{\by+\ex}\tau^Z_{\by-\ey}\tau^{Z\dagger}_{\by+\ey}.\nonumber
\end{eqnarray}
It is worth emphasizing that the gauged Hamiltonian contains the  terms of the toric code with dipole symmetry~\eqref{89}. The symmetry of the gauged Hamiltonian includes the following loop operators along~$x$-direction:
\begin{eqnarray}
    \xi^Z_{1}\vcentcolon=\prod_{\hx=1}^{L_x}\tau^Z_{(\hx,0)},\quad \xi^Z_2\vcentcolon=\prod_{\hx=1}^{L_x}\mu^Z_{(\hx+\frac{1}{2},\frac{1}{2})},\quad \xi^Z_{x,y}\vcentcolon=\prod_{\hx=1}^{L_x}\left[\left(\tau^{Z}_{(\hx,0)}\right)^{\hx}\right]^{\alpha_x},\label{loops9}
\end{eqnarray}
which are nothing but the modulated $1$-form symmetries.
When $L_x=0~\text{mod}~N$, the loops become identical to the ones in~\eqref{loops} with the relations~\eqref{dual94}, forming the dual dipole algebra. One can similarly show that the gauged model has modulated $1$-form symmetries in the $y$-direction, generated by noncontractible loop operators in the $y$-direction.
\subsubsection{Gauging $1$-form symmetry}\label{33}
Now we turn to gauging $\mathbb Z_N$ 1-form global symmetry of the model~\eqref{01spin}. To this end, we accommodate extended Hilbert space on each site of the lattice whose Pauli operator is denoted by $\widehat{\lambda}^{X/Z}_{\br}$. The Gauss's law is given by (Fig.~\ref{gauss11})
\begin{eqnarray}
    \widehat{\lambda}^Z_{\bx-\ex}Z_{\bx}  \widehat{\lambda}^{Z\dagger}_{\bx-\ex}=1,\quad  \widehat{\lambda}^Z_{\by-\ey}Z_{\by}  \widehat{\lambda}^{Z\dagger}_{\by-\ey}=1.\label{gauss4}
\end{eqnarray}
Intuition behind the Gauss's law term is that one decomposes the $1$-form operators~\eqref{1_form} into local pieces, in the same manner as the one in the previous subsection. \par
We modify the ``star operator" so that it commutes with the Gauss's law~\eqref{gauss4}:
\begin{eqnarray}
    X^{\dagger}_{\br-\ex}X_{\br+\ex}X^{\dagger}_{\br-\ey}X_{\br+\ey}\to X^{\dagger}_{\br-\ex}X_{\br+\ex}\widehat{\lambda}^{X\dagger}_{\br}X^{\dagger}_{\br-\ey}X_{\br+\ey}\label{gauss5}
\end{eqnarray}
Rewriting the right hand side of~\eqref{gauss5} as $\lambda^X_{\br}$, viz
\begin{eqnarray}
    \lambda^X_{\br}\vcentcolon=X^{\dagger}_{\br-\ex}X_{\br+\ex}\widehat{\lambda}^{X\dagger}_{\br}X^{\dagger}_{\br-\ey}X_{\br+\ey},
\end{eqnarray}
jointly with $\lambda^Z_{\br}\vcentcolon=\widehat{\lambda}^Z_{\br}$, 
the gauged Hamiltonian reads as
\begin{eqnarray}
    \Tilde{H}_{2D,1}=&-&J_x\sum_{\br}\left[\lambda^Z_{\br-\exx}\left(\lambda^{Z\dagger}_{\br}\right)^2\lambda^Z_{\br+\exx}+\lambda^Z_{\br}\lambda^{Z\dagger}_{\br+\exx}\lambda^{Z\dagger}_{\br+\eyy}\lambda^{Z}_{\br+\exx+\eyy}\right]\nonumber\\&-&J_y\sum_{\br}\left[\lambda^Z_{\br-\eyy}\left(\lambda^{Z\dagger}_{\br}\right)^2\lambda^Z_{\br+\eyy}+\lambda^Z_{\br}\lambda^{Z\dagger}_{\br+\exx}\lambda^{Z\dagger}_{\br+\eyy}\lambda^{Z}_{\br+\exx+\eyy}\right]\nonumber\\
    &-&J_G\sum_{\br}\lambda^X_{\br}+h.c..\label{gauged3}
\end{eqnarray}
This  Hamiltonian respects $0$-form modulated symmetry. To wit, the model commutes with the following operators:
\begin{eqnarray}
Q_{2D:0}=\prod_{\hx=1}^{L_x}\prod_{\hy=1}^{L_y}\lambda^X_{\br},\quad Q_{2D:x}=\left[\prod_{\hx=1}^{L_x}\prod_{\hy=1}^{L_y}(\lambda^X_{\br})^{\hx}\right]^{\alpha_x},\quad Q_{2D:y}=\left[\prod_{\hx=1}^{L_x}\prod_{\hy=1}^{L_y}(\lambda^X_{\br})^{\hy}\right]^{\alpha_y}.\label{51}
\end{eqnarray}
In particular, when $L_x=L_y=0~\text{mod}~ N$, it is identical to~\eqref{algebra}, forming the 0-form dipole algebra in~\eqref{algebra2}.\par

\subsection{Summary for 2D and comment on general dimensions}\label{3.3}
To recap the argument in this section, we study two 2D spin models with LSM-type anomalies, indicated by nontrivial commutation relations with dependence on system size. 
In the first model~\eqref{spin001} with two $0$-form global symmetries, 
gauging one of them yields a new type of modulated symmetry,  characterized by unusual dipole algebra containing $0$-form dipole and standard $1$-form symmetries. In the second model~\eqref{01spin} with two $0$-form and one $1$-form symmetries, gauging two $0$-form global symmetries yields modulated~$1$-form symmetry whereas gauging~$1$-form symmetry leads to modulated~$0$-form symmetry. \par 

It is widely known that when gauging $p$-form symmetry in $(d+1)$ spacetime dimension, one obtains dual $(d-1-p)$-form symmetry~\cite{gaiotto2015generalized}. Based on this fact, jointly with our results, we conjecture that in an anomalous spin system in $(d+1)$ spacetime dimension, where there are $p$- and $q$-form global symmetries ($0\leq p,q\leq d$, $0\leq p+q\leq d-1$) with nontrivial commutation relation depending on $L^{d-p-q}$~\footnote{Here, $L$ denotes linear system size of the system.}, gauging $p$-form symmetry gives a modulated symmetry in the dual theory, characterized by dipole algebra involving~$(d-1-p)$-form and $q$-form dipole symmetry. Further, hierarchical structure of the dipole algebra is formed in such a way that $q$-form symmetry is located at the higher hierarchy than the $(d-1-p)$-form symmetry. In other words, acting a translational operator on $q$-form symmetry produces $(d-1-p)$-form symmetry. 
To support the validness of this observation, we demonstrate an anomalous spin model defined in 3D in App.~\ref{sec4}
and obtain new modulated symmetries from gauging one of the global symmetries. Also, we give an alternative interpretation of our results in the field theoretical description in Sec.~\ref{5_1}, allowing us to understand the emergence of the modulated symmetries systematically.

\section{Field theoretical interpretation}\label{5_1}
It is often the case that an anomaly can be described by a topological action comprised of background gauge fields associated with global symmetries~\cite{Kapustin2014}. 
In this section, we give field theoretical interpretations of our results.
It turns out that the emergence of the modulated symmetries can be described in the similar manner as the generation of the higher groups~\cite{Cordova:2018cvg}.
To this end, we review 
gauge fields associated with the dipole symmetry and discuss 
how dipole symmetries are emerged in view of anomaly inflows and higher groups.

\subsection{Gauge fields associated with $p$-form dipole symmetries}\label{sec:51}
We first review gauge fields of $p$-form dipole symmetries~\cite{hirono2022symmetry,2023foliated,anomaly_2024}. 
Suppose we have a theory in $(d+1)$-spacetime dimension with conserved $p$-form charges associated with U(1) \textit{global} and \textit{dipole} symmetries, defined on $(d-p)$-dimensional spatial submanifold,~$\Sigma_{d-p}$.\footnote{%
Here we take $0\leq p\leq d$. The $p=0$ case corresponds to the ordinary global symmetry.}
We denote the global charge by~$Q[\Sigma_{d-p}]$ and dipole charge by~$Q_{I}[\Sigma_{-p}]$, where the index~$I =1,\cdots ,d$ denotes the dipole degrees of freedom in the~$I$-th spatial direction.~\footnote{We interchangeably represent the spatial direction $I=1,2,3,\cdots,$ as $I=x,y,z,\cdots,$~depending on the context.}\par
While the global charge $Q$ follows the relation
\begin{equation}
    [iP_I,Q]=0,\label{df}
\end{equation}
implying it is homogeneous in space, 
the dipole charges satisfy the following relation:
\begin{equation}
    [iP_I,Q_J]=\delta_{IJ}Q.\label{eq:re2}
\end{equation}
An intuitive picture is  considering the global and dipole charges densities $\rho$, $x_J\rho$ (where~$\rho$ denotes the density of the U(1) charge, and $x_J$ as the $J$-th spatial coordinate) under  translation in the $I$-th direction~($I,J=1,\cdots,d$)~\cite{2023foliated}. 
For instance, if we translate by a constant $\Delta x_I$ in the $I$-th~direction, then the change of dipole moment  gives $(x_I+\Delta x_I)\rho-x_I\rho=(\Delta x_I)\rho$, corresponding to the nontrivial commutation relation between the transnational operator and the dipole charge operator.  \par
We write the charges $Q$ and $Q_I$ via integral expression using the $(p+1)$-form conserved currents as
\begin{equation*}
   Q[\Sigma_{d-p}] =\int_{\Sigma_{d-p}}*j^{(p+1)},\quad  Q_I [\Sigma_{d-p}]=\int_{\Sigma_{d-p}}*K_I^{(p+1)} .
\end{equation*}
In order to satisfy the dipole algebra~\eqref{eq:re2}, we require that 
 \begin{equation}
    *K_I^{(p+1)}=*k_I^{(p+1)}-x_I*j^{(p+1)}\quad (I=1,\cdots,d)\label{dd}
\end{equation}
with $k_I\fl{p+1}$ being a local (non-conserved) current. 
Subsequently, we gauge the symmetries by
introducing~U(1) $(p+1)$-form gauge fields~$a\fl {p+1}$,~$A\fii{p+1}$ and minimally coupling them to the local currents\footnote{%
As discussed in~\cite{2023foliated}, one regards the gauge group as U(1), taking the fact that quantization condition of the dipole gauge field depends on the length of the dipole into consideration. We set such a length to be 1 throughout this section.}~\footnote{
Intuition behind this coupling is as follows: First, we introduce a gauge field $a\fl{p+1}$ to couple with the current $j^{(p+1)}$. Second, we introduce additional gauge fields $A\fii{p+1}$ to couple with the remnant part of the current in~\eqref{dd}, which is $k_I^{(p+1)}$.  }
\begin{equation}
    S_{cp}=\int_{V_{d+1}} \left( a\fl{p+1}\wedge *j\fl{p+1}+\sum_{I=1}^d A\fii{p+1}\wedge *k_I\fl{p+1} \right) ,
\end{equation}
where $V_{d+1}$ denotes the spacetime manifold.
A proper gauge transformation is required to gives rise to the conservation law of the higher form currents from the gauge invariance of this coupling term.
We illustrate this point in App.~\ref{u1} with the case of ordinary U(1)  symmetry.
It turns out that 
the following gauge transformation
\begin{equation}
    a\fl{p+1}\to a\fl{p+1}+d\Lambda\fl{p}+(-1)^{p}\sum_I\sigma\fii{p} \wde dx_I,\quad A\fii{p+1}\to A\fii{p+1}+d\sigma\fii{p},\label{gaugetr1}
\end{equation}
where $\Lambda\fl{p}$ and $\sigma\fii{p}$ denote the $p$-form gauge parameters, 
together with the gauge invariance of the coupling term $ S_{cp}$ yields the conservation law
\begin{equation*}
   d*j\fl{p+1}=0,\quad d(*k_I\fl{p+1}-x_I*j\fl{p+1})=d*K_I\fl{p+1}=0.
\end{equation*}
In what follows, $dx_I$ is interpreted as a $1$-form \textit{foliation field}~\cite{foliated2,foliated1}:
\begin{equation}
e^I \vcentcolon = dx_I,
\end{equation}
which is widely used in the context of fracton topological phases so that along the direction of foliation field, layers of co-dimension~$1$ submanifolds are stacked.
\par
For later purposes, we also define the \textit{dual dipole algebra} with an inverted hierarchy structure. 
Instead of~\eqref{df} and \eqref{eq:re2}, consider $d$ global charges $\widehat{Q}_I$ and one dipole charge $\widehat{Q}$ with relation
\begin{equation}
    [iP_I,\widehat{Q}_J]=0,\quad[iP_I,\widehat{Q}]=-\widehat{Q}_I.\label{e}
\end{equation}
Analogous to the argument below~\eqref{eq:re2},
this relation can be intuitively understood by acting translation on the dipole and $d$ global charge density  $\widehat{\eta}=-\sum_{I=1}^dx_I\rho_I$, and $\rho_I$. 
For example, by shifting the dipole~$\widehat{\eta}$ in the $I$-th direction, one obtains the second relation in~\eqref{e}.
Following the similar argument presented around \eqref{dd}-\eqref{gaugetr1}, we define gauge fields associated with the global and dipole charges~\eqref{e} as $b\fii{p+1}$ and $B\fl{p+1}$ respectively  with the following gauge transformations:
\begin{eqnarray}
   b\fii{p+1}\to b\fii{p+1}+d\widehat{\chi}\fii{p}-(-1)^{p}\widehat{\sigma}\fl{p} \wde e^I,\quad B\fl {p+1}\to B\fl {p+1}+d\widehat{\sigma}\fl{p},\label{ddd0}
\end{eqnarray}
with $p$-form gauge parameters $\widehat{\chi}\fii{p}$ and $\widehat{\sigma}\fl{p}$.\par
The dipole algebra \eqref{eq:re2} is related to the dual one \eqref{e} by inverting the hierarchy structure of the algebra. 
Stated symbolically, 
\begin{equation}
    \begin{Bmatrix}
Q_1 & Q_2 & \cdots&Q_d\\
& Q &&
\end{Bmatrix}
\quad \leftrightarrow\quad 
    \begin{Bmatrix}
    & \widehat{Q} &&\\
\widehat{Q}_1 & \widehat{Q}_2 & \cdots&\widehat{Q}_d
\end{Bmatrix}.\label{triangle}
\end{equation}
These algebras put different mobility constraints on charges. 
In the case of the dipole algebra~\eqref{eq:re2}, a single charge is immobile as dipole moment is conserved in any spatial direction. 
On the contrary, in the case of the dual dipole algebra~\eqref{e}, which consists of $d$ charges (labeled by $I=1,\cdots,d$) and one dipole, the $I$-th charge, $\widehat{Q}_I$, is mobile in the direction perpendicular to the $I$-th direction, yet it is immobile in the~$I$-th direction. 
\par

The gauge invariant fluxes are introduced as 
\begin{eqnarray}
    f_a^{(p+1)}\vcentcolon=da\fl{p+1}+(-1)^{p+1}\sum_{I=1}^dA\fii{p+1}\wedge e^I,\quad F_A^{I(p+1)}\vcentcolon=d\Tilde{A}\fii{p+1}\\
    f_b^{I(p+1)}\vcentcolon=d{b}\fii{p+1}+(-1)^pB\fl{p+1}\wedge e^I,\quad F_B^{I(p+1)}\vcentcolon= dB\fii{p+1}.
\end{eqnarray}
The flatness conditions of the gauge fields are given by
\begin{eqnarray*}
   f_a^{(p+1)}=0,\quad F_A^{I(p+1)}=0,\\
    f_b^{I(p+1)}=0,\quad F_B^{(p+1)}=0
\end{eqnarray*}
which are rewritten as 
\begin{eqnarray}
    da\fl{p+1}=(-1)^{p}\sum_{I=1}^d{A}\fii{p+1}\wedge e^I,\quad d{A}\fii{p+1}=0\label{95},\\
   db\fii{p+1}=(-1)^{p+1}{B}\fl{p+1}\wedge e^I,\quad d{B}\fl{p+1}=0.\label{96}
\end{eqnarray}
Although in this subsection we assume $U(1)$ symmetry for simplicity, the nontrivial flatness condition for gauge field of dipole symmetry can be generalized to $\mathbb{Z}_N$ gauge theory by Higgsing the $U(1)$ gauge theory by a charge $N$ field.

\subsection{$p$-form dipole symmetries from generalized LSM type anomaly}

\subsubsection{1D example}
In Sec.2.1.2, we reviewed how to obtain $\mathbb Z_N$ dipole symmetry algebra from a $\mathbb Z_N\times \mathbb Z_N$ symmetry with the LSM type commutation relation between symmetry generators
\begin{equation}
    U^{(0)}_ZU^{(0)}_X=\omega^LU^{(0)}_XU^{(0)}_Z,
\end{equation}
on a $\mathbb Z_N$-qudit spin chain. This can be shown from a field theory perspective. This LSM type anomaly is captured by a $(2+1)d$ $\mathbb Z_N\times \mathbb Z_N$ weak symmetry protected topological (SPT) phases~\cite{2024multipole,Pace:2025hpb,PhysRevLett.98.106803,You:2016unp}
\begin{equation}\label{eq:3dbulkanomaly}
    \frac{iN}{2\pi}\int_{M_3}G^{(1)}\wedge H^{(1)}\wedge e^x,
\end{equation}
protected by internal $\mathbb Z_N\times \mathbb Z_N$ symmetry and lattice translation symmetry. Weak SPTs are invertible foliated field theories, serving as anomaly-inflow bulk theories for the boundary LSM anomaly between internal symmetry and lattice translation symmetry~\cite{Cheng:2015kce,Cheng:2018lti}. Here $G^{(1)},~H^{(1)}$ denote 1-form background $\mathbb Z_N$ gauge fields.~\footnote{In the gauging process, we use the capital letters $G,H$ for background $\mathbb Z_N$ gauge field and $g,h$ for dynamical gauge field. This should not be confused with the convention we used to derive gauge field for dipole symmetry, where we use the capital letters $A,B$ for  gauge field coupled to currents of dipole symmetry  and $a,b$ for gauge field coupled to current of ordinary symmetry.} 
Now we gauge the $\mathbb Z_N$ symmetry with gauge field $G^{(1)}$ by promoting it to dynamical gauge field $g^{(1)}$ and coupling to the dual background gauge field $\tilde{G}^{(1)}$. The gauge symmetry should be free of anomaly, which means we need to couple
\begin{equation}
    \frac{iN}{2\pi}\int_{\partial M_3} g^{(1)}\wedge \tilde{G}^{(1)}=\frac{iN}{2\pi}\int_{M_3} d(g^{(1)}\wedge \tilde{G}^{(1)})=\frac{iN}{2\pi}\int_{M_3} -g^{(1)}\wedge d\tilde{G}^{(1)},
\end{equation}
at the boundary to cancel the bulk anomaly term \eqref{eq:3dbulkanomaly}. We used the flatness condition $dg^{(1)}=0$ for the~$\mathbb Z_N$ gauge field $g^{(1)}$ in the last equality. Therefore, we have
\begin{equation}
    g^{(1)}\wedge H^{(1)}\wedge e^x -g^{(1)}\wedge d \tilde{G}^{(1)}=0.
\end{equation}
This leads to the modified flatness condition for the dual gauge field
\begin{equation}
    d\tilde{G}^{(1)}=H^{(1)}\wedge e^x.
\end{equation}
Together with $dH^{(1)}=0$, the gauged theory exhibits the same flatness condition as $\eqref{95}$ for the case of the 0-form dipole symmetry.~\footnote{We thank Linhao Li for 
helpful discussion on this point.} 

\subsubsection{General dimensions}
We will generalize the argument in $(1+1)d$ to general dimensions. We start from a~$(d+1)$-dimensional theory with LSM type anomaly between internal symmetries and lattice translation symmetry. By gauging one of the internal symmetries, we expect to get a dual modulated symmetry with gauge fields following a dipole-like flatness condition.

For simplicity, consider 
one $p$-form and $d$ copies of
$(d-p-1)$-form $\mathbb Z_N$~global symmetries 
with background gauge fields $G\fl{p+1}$ and $H\fii{d-p}$, where $I=1,\cdots,d$ distinguishes different $(d-p-1)$-form symmetries. 
Assuming these symmetries have 
LSM type anomaly described by
\begin{eqnarray}
    \boxed{ 
    S=\sum_{I=1}^d\frac{iN}{2\pi}\int_{M_{d+2}}G^{(p+1)}\wedge H^{I(d-p)}\wedge e^I\label{general},
}
\end{eqnarray}
where the foliation field~$e^I$ is regarded as the gauge field associated with translation symmetry in the~$I$-th direction.
This is a natural generalization of the weak SPT in $(2+1)d$. 
{Physically, the terms in~\eqref{general} describe SPT phases protected by $\mathbb{Z}_N$ $p$-form and $(d-p-1)$-form symmetries stacked along in the~$I$-th direction. An explicit lattice model of~\eqref{general} in $(3+1)d$ was constructed in App.~B of~\cite{anomaly_2024}.}
For the anomalous boundary theory on the $d$-dimensional square lattice, this anomaly term is reflected by $d$ copies of commutation relations with dependence on linear system size on each direction, for example,
\begin{equation}
    U_G^{(p)}(\Sigma_{d-p})U_H^{I,(d-p-1)}(\Sigma^I_{p+1})=\omega^{L_I}U_H^{I,(d-p-1)}(\Sigma^I_{p+1})U_G^{(p)}(\Sigma_{d-p}),\quad I=1,\cdots, d,
\end{equation}
where $U_G^{(p)}(\Sigma_{d-p})$ and $U_H^{I,(d-p-1)}(\Sigma^I_{p+1})$ represent $\mathbb{Z}_N$ $p$-form and $(d-p-1)$-form symmetry operators which have support on $\Sigma_{d-p}$ and  $\Sigma^I_{p+1}$, respectively. Also, the intersection of $\Sigma^I_{p+1}$ and~$\Sigma_{d-p}$ is a line in the $I$-th direction. \par

Now we are in a good place to perform
gauging the $p$-form symmetry by making it  dynamical and coupling it to the dual background field through
\begin{equation}
    \frac{iN}{2\pi}\int_{\partial M_{d+2}} g^{(p+1)}\wedge \tilde{G}^{(d-p)}=\frac{iN}{2\pi}\int_{M_{d+2}} d(g^{(p+1)}\wedge \tilde{G}^{(d-p)})=\frac{iN}{2\pi}\int_{M_{d+2}} (-1)^{p+1}\ g^{(p+1)}\wedge d\tilde{G}^{(d-p)}.
\end{equation}
This term should cancel the bulk anomaly by
\begin{equation}
    \sum_{I=1}^d g^{(p+1)}\wedge H^{I(d-p)}\wedge e^I+(-1)^{p+1}\ g^{(p+1)}\wedge d\tilde{G}^{(d-p)}=0,
\end{equation}
which leads to the modified flatness condition 
\begin{equation}
    d\tilde{G}^{(d-p)}=(-1)^{p}\sum_{I=1}^d H^{I(d-p)}\wedge e^I,
\end{equation}
for gauge fields of $(d-p-1)$-form dipole symmetry.

This derivation closely parallels the one studied in the context of higher group~\cite{Tachikawa:2017gyf,Cordova:2018cvg}: for a theory in~$(d+1)$-dimension with two global symmetries with mixed~\tht~anomaly, gauging one of the global symmetries leads to the dual theory with a nontrivial extension between dual symmetries (which forms a higher group structure). During the gauging procedure, we also trivialize the $(d+2)$-dimensional dependence from the anomaly counterterm by modifying the cocycle (flatness) condition of the gauge fields.  
We give a lightning review of this problem in App.~\ref{app2}.\par
By the same logic, in the same setting with LSM type anomaly~\eqref{general}, if we gauge $(d-p-1)$-form symmetries, one can trivialize the $(d+2)$-dimensional dependence by imposing
the following flatness condition of the gauge field:
\begin{eqnarray}
    d\tilde{H}\fii{p+1}=(-1)^{p(d-p)+1}G\fl{p+1}\wedge e^I,\quad dG\fl{p+1}=0,
\end{eqnarray}
where $\tilde{H}\fii{p+1}$ denotes gauge field of the $p$-form dual symmetry. Up to the minus sign,  
this flatness condition corresponds to the one for the gauge fields associated with $p$-form dual dipole symmetry in~\eqref{96}.  
\par
In summary, we introduce a $(d+1)$-dimensional theory where there are
$p$-form and $(d-p-1)$-form $\mathbb Z_N$~global symmetries with LSM type anomaly~\eqref{general}. 
Gauging either of the global symmetry leads to a dual modulated symmetry. 
This is a natural generalization from 
the previous study of modulated symmetries in 1D~\cite{Seifnashri:2023dpa,Aksoy:2023hve} and 
in 2D~\cite{Ebisu:2024eew}. In the case of 1D, 
gauging one of $0$-form symmetries with an LSM type anomaly described by the inflow term~\eqref{general} with $p=0$ and $d=1$  gives rise to dipole symmetry with $0$-form dipole algebra. 
In the case of 2D, for a theory with two $0$-form and one $1$-form symmetries with an LSM type anomaly described by the inflow term~\eqref{general} with $p=0,d=2$, gauging two $0$-form symmetries leads to $1$-form dipole symmetry with dual dipole algebra. The corresponding lattice model is in Sec.~\ref{32} [see~\eqref{dual94}]. Likewise, if we instead gauge~$1$-form symmetry, we would end up with $0$-form dipole symmetry with dipole algebra, whose lattice realization is discussed in Sec.~\ref{33} [see also~\eqref{algebra2}].

\subsection{Dipole symmetries involving different forms from generalized LSM type anomaly}

In previous subsections, we review the properties of gauge field for $p$-form dipole symmetry. We show that the modified flatness condition including foliation fields can be obtained by gauging a theory with the LSM type anomaly, which is captured by a weak SPT phase in one higher dimension. The weak SPT is described by an invertible foliation field theory with one layer of foliation, reflecting the anomalous phase with linear system size dependence on the lattice. 

However, we have discovered new types of modulated symmetries mixing between symmetries of different forms in lattice models. These new symmetries can be obtained from a generalized LSM type anomaly with anomalous phase depending on the area of the system. In this subsection, we initiated the study of gauge fields for these novel symmetries in the field theory perspective for examples in $(2+1)d$. We also show that by gauging from a invertible foliated field theory with \textit{two} layers of foliation, which captures the generalized LSM type anomaly, the properties of gauge field associated with the new modulated symmetries emerges naturally. 

\begin{figure}
    \begin{center}
  \includegraphics[width=0.3\textwidth]{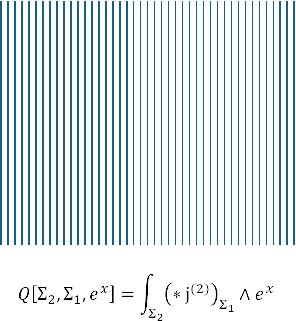}
\end{center}
 \caption{Visual illustration of the charge $  Q[\Sigma_2,\Sigma_1^y,e^x]$, which is composed of 
 stack of the currents $*j\fl2$ located on $\Sigma_1^y$ (blue lines on the right hand side) along the $x$-direction.   }\label{tr10}
 \end{figure} 
Consider a theory in $(2+1)d$ with one $1$-form and one $0$-form U(1)~symmetries whose corresponding conserved currents are given by $j^{(2)}$ and~$K\fl1$.
From these currents, one could construct charges as
\begin{eqnarray*}
    Q=\int _{\Sigma_1}*j\fl2,\quad Q^\prime=\int_{\Sigma_2}*K\fl1.
\end{eqnarray*}
Instead of doing it, 
we would like to have an algebraic relation between charges with different forms via translational operators, in analogous to the previous discussion on the dipole algebra. However, space dimensions where these charges are defined are different. To circumvent this issue, we propose 
 the following charges:
\begin{eqnarray}
    Q[\Sigma_2,\Sigma_1,e^x]=\int_{\Sigma_2}\left(*j\fl2\right)_{\Sigma_1}\wedge e^x,\quad Q^\prime[\Sigma_2]=\int_{\Sigma_2}*K\fl1\label{charges1}
\end{eqnarray}
While the second term is 
the standard expression of the charge for 
 $0$-form symmetry, 
 the first terms includes higher form current $\left(*j\fl2\right)_{\Sigma_1}$, located on one dimensional rigid slices~$\Sigma_1$ along the $y$-direction and stacked along the other direction through the foliation field. See also Fig.~\ref{tr10} for illustration. This is consistent with the foliated 1-form charges in the lattice dipole algebra~\eqref{dipole01}.
Here, we specify the foliation field~$e^x$ in the first term of~\eqref{charges1}, hence, the way $0$-form charge is defined depends on the manifold $\Sigma_1$ and the foliation field.
In what follows, we retain such manifold and foliation field dependence of the charge (More explicitly, we write the two charges in~\eqref{charges1} as $Q[\Sigma_1,e^x]$ and $Q^\prime$.). 
To proceed, we impose the conditions on charges:
\begin{eqnarray}
    [iP_I,Q[\Sigma_1,e^x]]=0~(I=x,y),\quad [iP_y,Q^{\prime}]=Q[\Sigma_1,e^x],\quad [iP_x,Q^{\prime}]=0,\label{0,1}
\end{eqnarray}
which are generalization of dipole algebra involving different form of symmetries. 
In the following, we introduce gauge fields corresponding to the dipole algebra~\eqref{0,1}. To do so, in analogy to the discussion in the previous subsection, we rewrite the current $*K^{(1)}$ as
\begin{eqnarray}
    *K\fl1=*k\fl1-y\left(*j\fl2\right)\wedge e^x\label{102}
\end{eqnarray}
with $k^{(1)}$ being non-conserved current. A simple calculation gives~\eqref{0,1} from~\eqref{102}. Introducing $2$-form and $1$-form gauge field, $a\fl2$ and $A\fl1$, 
we think of the following coupling term:
\begin{eqnarray}
    \widehat{S}=\int_{V_3}\left( a\fl2\wedge *j\fl2+A\fl1\wedge *k\fl1\right)\label{cp2}
\end{eqnarray}
If we demand the following gauge transformation~\footnote{In the rest of this section, $\lambda\fl p$ and$\Lambda\fl q$ denote $p$-from and $q$-form gauge parameters.  }
\begin{eqnarray}
    a\fl2\to a\fl2+d\lambda\fl1+\Lambda\fl0 e^x\wedge e^y,\quad A\fl1\to A\fl1+d \Lambda\fl0,
\end{eqnarray}
then gauge invariance of the coupling term~\eqref{cp2} leads to $d*j\fl2=d*K\fl1=0$.
Defining gauge invariant fluxes as 
\begin{eqnarray}
    f\fl3_a=da\fl2-A\fl1\wedge e^x\wedge e^y,\quad F_A\fl2=dA\fl1,
\end{eqnarray}
the flatness condition of the gauge fields reads $f_a\fl3=F_A\fl2=0$, viz,
\begin{eqnarray}
    da\fl2=A\fl1\wedge e^x\wedge e^y,\quad dA\fl1=0.\label{flat2}
\end{eqnarray}
In summary so far, we gauge dipole symmetry involving two global symmetries with different forms. The gauge fields associated with such a symmetry are subject to the flatness condition given by~\eqref{flat2}. Note that had we choose one foliation field $e^y$ instead of $e^x$ and define charges as
\begin{eqnarray}
       Q[\Sigma_2,\Sigma_1,e^y]=\int_{\Sigma_2}\left(*j\fl2\right)_{\Sigma_1}\wedge e^y,\quad Q^\prime[\Sigma_2]=\int_{\Sigma_2}*K^{(1)}
\end{eqnarray}
and think of the following dipole algebra
\begin{eqnarray}
    [iP_I,Q[\Sigma_1,e^y]]=0~(I=x,y),\quad [iP_x,Q^{\prime}]=Q[\Sigma_1,e^y],\quad [iP_y,Q^{\prime}]=0\label{0,12},
\end{eqnarray}
we would arrive at the same flatness condition~\eqref{flat2} when gauging dipole symmetry. 

Now we derive this modified flatness condition for gauge field corresponding to the dipole symmetry~\eqref{0,1} from the generalized ~\tht~ anomaly. As discussed in~\ref{00}, the anomalous phase for two $\mathbb Z_N$ $0$-form global symmetries $(2+1)$d depends on the area of the system. This implies that this anomalous theory lives on the boundary of a 2-foliated invertible theory~\footnote{Field theories with two foliation fields were discussed in~\cite{Hsin:2024eyg} in a different context. In our case, such two foliation fields are introduced to discuss the LSM anomaly between two global and translational symmetries in the $x$- and $y$-direction.}
\begin{eqnarray}
    \boxed{S=\frac{iN}{2\pi}\int_{M_{4}}G^{(1)}\wedge H^{(1)}\wedge e^x\wedge e^y.}
\end{eqnarray}
where $G^{(1)}, H^{(1)}$ are background gauge fields for these two $\mathbb Z_N$ symmetries.
When we gauge the $\mathbb Z_N$ symmetry with gauge field $G\fl1$, we obtain a dual $1$-form symmetry with gauge field $\tilde{G}\fl{2}$. To cancel the bulk anomaly term and trivialize the four spacetime dimensional dependence, 
we follow similar discussion in the previous subsections, and obtain the correct modified flatness condition
\begin{eqnarray}
    d\tilde{G}\fl{2}=H\fl{1}\wedge e^x\wedge e^y,\quad dH\fl1=0,
\end{eqnarray}
which is identical to~\eqref{flat2}.


\begin{table}[h]
\begin{center}
\begin{tabular}{ |c|c|c| c|c|} 
 \hline
 Dim & $p$- and $q$-form sym & Bulk invertible theory & Dipole sym & Dipole sym \\
  &   & for LSM anomaly &  by gauging $p$&  by gauging $q$\\\hline\hline
 1D & $(p,q)=(0,0)$ & $\int_{M_{3}}G^{(1)}\wedge H^{(1)}\wedge e^x$& $0$-form~$\xrightarrow[]{T}$~$0$-form & $0$-form~$\xrightarrow[]{T}$~$0$-form \\ 
 Sec.~\ref{1dmd} & $G^{(1)}$, $H^{(1)}$ & &  &\\
 \hline
2D & $(p,q)=(0,0)$ & $\int_{M_{4}}G^{(1)}\wedge H^{(1)}\wedge e^x\wedge e^y$& $0$-form~$\xrightarrow[]{T}$~$1$-form & $0$-form~$\xrightarrow[]{T}$~$1$-form \\ 
Sec.~\ref{00} & $G^{(1)}$, $H^{(1)}$ & & & 
\\\hline
2D & $(p,q)=(0,1)$&$\int_{M_{4}}G^{I,(1)}\wedge H^{(2)}\wedge e^{I}$
&$1$-form~$\xrightarrow[]{T}$~$1$-form & $0$-form~$\xrightarrow[]{T}$~$0$-form \\
Sec.~\ref{32} & $G\fii1$($I=1,2$), $H^{(2)}$& 
& & \\
\hline
3D & $(p,q)=(0,1)$& $\int_{M_{5}}G^{(1),I}\wedge H^{(2)}\wedge e^{J}\wedge e^K$
&$1$-form~$\xrightarrow[]{T}$~$2$-form & $0$-form~$\xrightarrow[]{T}$~$1$-form \\
App.~\ref{sec4} & $G^{(1),I}$($I=1,2,3$), $H^{(2)}$& $(I\neq J\neq K)$
& &\\
 \hline
 \end{tabular} 
    \caption{Summary of this subsection. We think of a theory with $p$- and $q$-form symmetries in $(d+1)$ spacetime dimension whose corresponding gauge fields are $G\fl{p+1}$ and $H\fl{q+1}$, respectively, with the anomaly described by the third column. By gauging $p$- or $q$-form symmetry, one obtains dipole symmetry, described by a dipole algebra consisting of different form of symmetries. In the second line of the first column, we refer to the section where the corresponding lattice model is discussed. By gauging one of the global symmetries, we obtain dipole symmetry, described by a dipole algebra, consisting of $p^\prime$-form and $q^\prime$-form symmetries, the latter of which is generated by acting a translational operator (represented by ``$T$" in the fourth and fifth column ) on the former.  }
    \label{222}
\end{center}
\end{table}
\subsection{Summary of this section}\label{5.5}
By investigating dipole algebra and gauge fields associated with them in the field perspective, we elucidate that emergence of~$\mathbb Z_N$~dipole algebra can be interpreted as gauging one of the global symmetries with an anomalous system involving foliation field(s). The emergence of such dipole symmetry 
corresponds to the our $\mathbb{Z}_N$ spin model on a discrete lattice. 
In App.~\ref{3dth}, we discuss the anomaly inflow terms involving two foliation fields that comply with our 3D lattice model given in App.~\ref{sec4}.  
We summarize the consideration given in this section in Table.~\ref{222} 
\section{Discussion}\label{sec5}
To address the question ``how does modulated symmetry emerge?", 
in this work, we have presented explicit lattice models defined in two and three spatial dimension, 
possessing global symmetries with the LSM-like anomaly -- global symmetries exhibiting nontrivial commutation relations depending on the system size. We elucidate that depending on the form of the global symmetries, there are various dipole symmetries: 
Suppose the model has $p$-form and $q$-form global symmetries
in $d$ spatial dimension ($0\leq p,q\leq d$), with an anomaly in the sense that commutation relation between  $p$-form and $q$-form symmetry operators depend on system size $L^{d-(p+q)}$, or put simply the two global symmetries and the lattice translation have the LSM anomaly. Then, 
gauging $p$-form symmetry yields a dipole symmetry, described by dipole algebra consisting of emergent $[d-(p+1)]$-form and $q$-form symmetries. More explicitly, the dipole algebra is formed in a such a way that acting a translational operator on the $q$-form symmetry generates $[d-(p+1)]$-form symmetry. 
We give field theoretical argument to understand the relation between the emergence of the modulated symmetry and the LSM anomaly. 
Our work provides a new perspective of the emergence of modulated symmetries in a concrete quantum lattice system with anomaly, making better understanding of these exotic symmetries, especially the ones in spatial dimension more than one. 
We emphasize that our 
method to get modulated symmetry holds generally, beyond the exactly solvable models we provide in this paper.
\par
We close this section by giving a few future directions. In this paper, we discuss the relation between modulated symmetries and the LSM type anomaly involving translational symmetries. One would naively wonders whether it can be generalized to a system with another type of the LSM anomaly, associated with crystalline symmetries, such as rotation and reflection. We have a speculation that given two global symmetries, whose gauge fields are $\alpha\fl p$ and $\beta\fl p$, we start with an anomaly term which has the form of
\begin{eqnarray*}
   \int_{M_{d+2}}\alpha\fl p\wedge \beta \fl q \wedge C(e^I,\omega,\cdots,),
\end{eqnarray*}
where $C$ describes a gauge field associated with crystalline symmetry, such as translation and rotation, whose gauge fields are denoted as $e^I$ and $\omega$, respectively, 
more general modulated symmetries are ubiquitously generated via gauging one of the global symmetries. Studying emergence of the modulated symmetries in this more broad perspective would deepen our understanding of the concept of symmetries.
In this paper, we focus on a theory with two global symmetries. 
One could extend the analysis to the case of lager number of internal symmetries. In such a case, one would expect higher rank of multipole symmetries, such as quadrupole symmetry~\cite{ebisu2023symmetric,Saito:2025qrp}. Also, studying anomaly counter term involving more general foliation structure~\cite{Ebisu:2024mbb} would contribute to better understanding of the modulated symmetries. 
\par
Also, it would be interesting to address how modulated symmetries that we have studied in this paper would influence on dynamics of a system. Since our model are described by spin systems on a lattice, one could study more practical aspect of the model, such as what is the behavior of the Hilbert space fragmentation or thermalization (See e.g.,~\cite{Miao:2025hvw}.). We hopefully come back to these issues in the future.
\section*{Acknowledgment}
We would like to thank 
D. Bulmash, M. Honda, Z. Jia, A. Kurebe, L. Li, 
T. Nakanishi, 
 A. Nayak, K. Shiozaki, P. Tanay,  T. Saito , P. M. Tam, 
 A. Ueda, S. Pace, A. Yosprakob, H. Watanabe for helpful discussion. This work is in part supported by JST CREST (Grant No.~JPMJCR24I3), Villum Fonden Grant no.~VIL60714,  
the Deutsche Forschungsgemeinschaft (DFG, German Research Foundation) under Germany’s Excellence Strategy—Cluster of Excellence Matter and Light for Quantum Computing (ML4Q) EXC 2004/1 -- 390534769 as well as within the CRC network TR 183 (Project Grant No. 277101999) as part of subproject B01. H.~E. acknowledges the hospitality of the
Institute for Theoretical Physics during the visit in Cologne, where part of this work was performed.

\appendix
\section{Emergence of modulated symmetry from 3D model with the LSM anomaly}\label{sec4}
In this appendix, we demonstrate our conjecture through a spin model on a cubic lattice with three $0$-form and one $1$-form symmetries with a generalized LSM type anomaly. This LSM anomaly combines two generalized features in 2D: (i)~It is the mixed \tht~ anomaly between lattice translation and internal symmetries of \textit{different} forms. (ii)~The anomalous phases in the commutation relations depend on the \textit{area} of the system $O(L^2)$. Via gauging one of global symmetries, we obtain novel modulated symmetries characterized by dipole algebra, involving $p$-form and $q$-form symmetries with $p\neq q$. \par

\subsection{Hamiltonian}
To start, we consider $\mathbb{Z}_N$ spin  on each link of a cubic lattice with ~$L_x\times L_y\times L_z$ sites and periodic boundary condition. We introduce the following Hamiltonian:
\begin{eqnarray}
    H_{3D}=-J_x\sum_{\bx}\left(Z^{\dagger}_{\bx+\exx}Z_{\bx}+Z^{\dagger}_{\bx+\eyy}Z_{\bx}+Z^{\dagger}_{\bx+\ezz}Z_{\bx}\right)-J_y\sum_{\by}\left(Z^{\dagger}_{\by+\exx}Z_{\bx}+Z^{\dagger}_{\by+\eyy}Z_{\bx}+Z^{\dagger}_{\by+\ezz}Z_{\bx}\right)\nonumber\\
    -J_z\sum_{\bz}\left(Z^{\dagger}_{\bz+\exx}Z_{\bx}+Z^{\dagger}_{\bz+\eyy}Z_{\bx}+Z^{\dagger}_{\bz+\ezz}Z_{\bx}\right)-J_p\sum_{\bp_{ab}}\mathcal{P}^X_{\bp_{ab}}-J_G\sum_{\br}\mathcal{Q}^Z_{\br}+h.c.,\label{52}
\end{eqnarray}
with 
\begin{eqnarray}
\mathcal{P}^X_{\bp_{ab}}\vcentcolon=X_{\mathbf{l}_a+\mathbf{e}_b}X^{\dagger}_{\mathbf{l}_a}X_{\mathbf{l}_b+\mathbf{e}_a}^{\dagger}X_{\mathbf{l}_b},\quad \mathcal{Q}^Z_{\br}\vcentcolon=X_{\br+\ex}X^{\dagger}_{\br-\ex}X_{\br+\ey}X^{\dagger}_{\br-\ey}X_{\br+\ez}X^{\dagger}_{\br-\ez}.\label{3dtoric}
\end{eqnarray}
\begin{figure}
\hspace{15mm}
         \begin{subfigure}[h]{0.55\textwidth}
       \centering
  \includegraphics[width=1.4\textwidth]{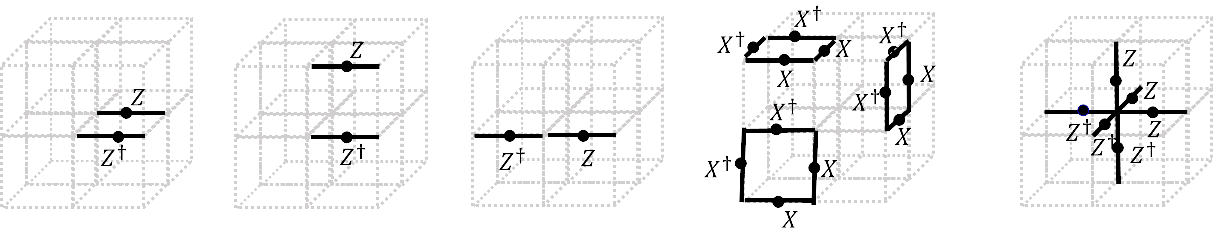}
         \caption{}\label{3D}
             \end{subfigure}
            
            \hspace{30mm}\begin{subfigure}[h]{0.55\textwidth}
            \centering
  \includegraphics[width=0.9\textwidth]{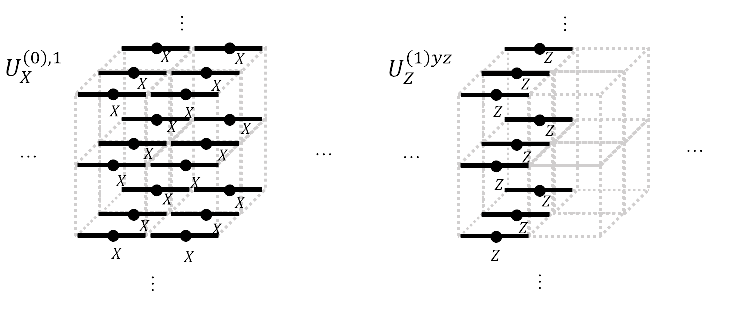}
         \caption{}\label{3D2}
             \end{subfigure}
                         
                         \hspace{30mm}
                \begin{subfigure}[h]{0.55\textwidth}
            \centering
  \includegraphics[width=0.9\textwidth]{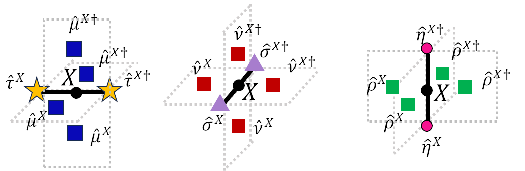}
         \caption{}\label{3D3}
             \end{subfigure}
            
 \caption{(a)~[The first three configurations]:~Spin coupling terms corresponding to the first three terms in~\eqref{52}. [The last two configurations]: terms given in~\eqref{3dtoric} that constitute 3D toric code. 
 (b)~One of $0$-form [$1$-form symmetry] operators given in~\eqref{global33}~[\eqref{membrane}] on the left [right] configuration. (c)~Gauss's laws defined in~\eqref{gauss6}.  }
 \end{figure}
Here, $\bp_{ab}$ with $ab=xy,yz,zx$ denotes coordinate of a plaquette on $ab$-plane, that is, $\bp_{xy}=(\hx+\frac{1}{2},\hy+\frac{1}{2},\hz)$, $\bp_{yz}=(\hx,\hy+\frac{1}{2},\hz+\frac{1}{2})$, $\bp_{zx}=(\hx+\frac{1}{2},\hy,\hz+\frac{1}{2})$.
The first three terms 
as well as the last two
in~\eqref{52} are portrayed in Fig.~\ref{3D}.
Note that the last two terms in~\eqref{52} describe the 3D $\mathbb{Z}_N$ toric code. In what follows, we set $J_G\to \infty$ to ensure the fluxless condition, that is, we focus on a state $\ket{\Omega}$ satisfying 
\begin{eqnarray}
\mathcal{Q}_{\br}^Z\ket{\Omega}=\ket{\Omega},\forall \br.\label{fluxless}
\end{eqnarray}
The model~\eqref{52} have the following three $\mathbb Z_N$ $0$-form global symmetries generated by
\begin{eqnarray}
    U^{(0),1}_X=\prod_{\hx=1}^{L_x}\prod_{\hy=1}^{L_y}\prod_{\hz=1}^{L_z}X_{\bx},\quad U^{(0),2}_X=\prod_{\hx=1}^{L_x}\prod_{\hy=1}^{L_y}\prod_{\hz=1}^{L_z}X_{\by},\quad U^{(0),3}_X=\prod_{\hx=1}^{L_x}\prod_{\hy=1}^{L_y}\prod_{\hz=1}^{L_z}X_{\bz}.\label{global33}
\end{eqnarray}
In addition, the model admits  one $\mathbb Z_N$ $1$-form symmetry, generated by the following noncontractible membrane operators
\begin{eqnarray}
    U^{(1),xy}_Z=\prod_{\hx=1}^{L_x}\prod_{\hy=1}^{L_y} Z_{(\hx,\hy,\hz+\frac{1}{2})},\quad U^{(1),yz}_Z=\prod_{\hy=1}^{L_y}\prod_{\hz=1}^{L_z} Z_{(\hx+\frac{1}{2},\hy,\hz)},\quad U^{(1),zx}_Z=\prod_{\hz=1}^{L_z}\prod_{\hx=1}^{L_x} Z_{(\hx,\hy+\frac{1}{2},\hz)}.\label{membrane}
\end{eqnarray}
Note that the membrane operators are topological: they only depend on the nontrivial homology of the lattice due to the fluxless condition~\eqref{fluxless}. 
These $0$-form and $1$-form global symmetries exhibit nontrivial commutation relations:
\begin{eqnarray}
   U^{(1),yz}_ZU^{(0),1}_X&=&\omega^{L_yL_z}U^{(0),1}_XU^{(1),yz}_Z\nonumber\\ U^{(1),zx}_ZU^{(0),2}_X&=&\omega^{L_zL_x}U^{(0),2}_XU^{(1),zx}_Z \nonumber\\
   U^{(1),xy}_ZU^{(0),3}_X&=&\omega^{L_xL_y}U^{(0),3}_XU^{(1),xy}_Z,
\end{eqnarray}
with anomalous phases depending on the area of the system $O(L^2)$.
\par
Based on our conjecture~(Sec.~\ref{3.3}), one could speculate what kind of modulated symmetries are generated in this system. 
Gauging three $0$-form symmetries yields dual $2$-form symmetries and the consequent dipole algebra mixes between $1$-form and $2$-form symmetries. On the other hand, after gauging the $1$-form symmetry, one obtains a dual $1$-form symmetry, leading to a dipole algebra involving $0$-form dipole symmetry and $1$-form ordinary symmetry. In the following subsections, we show that this speculation is correct by explicitly performing gauging in this lattice model. Further, a detailed field theoretical analysis on our model is given in App.~\ref{3dth}.

\subsection{Gauging $0$-form symmetry}
Let us first focus on gauging three $0$-form symmetries~\eqref{global33}. To this end, we introduce three $\mathbb{Z}_N$ spins on each node whose $X$ Pauli operators are denoted by $\widehat{\tau}_{\br}^X$, $\widehat{\sigma}_{\br}^X$, $\widehat{\eta}_{\br}^X$ with Pauli $Z$ operators being analogously defined. 
Further, we introduce three types of $\mathbb{Z}_N$ spins on plaquettes. Their Pauli $X$ operators are represented by $\widehat{\mu}^X_{\bp_{ab}}~(ab=xy,zx)$
, $\widehat{\nu}^X_{\bp_{cd}}~(cd=xy,yz)$, $\widehat{\rho}^X_{\bp_{ef}}~(ef=zx,yz)$. Pauli $Z$ operators are similarly defined. The Gauss's laws are
\begin{eqnarray}
        \widehat{\tau}^{X\dagger}_{\bx+\ex}\widehat{\mu}^{X\dagger}_{\bx+\ey}\widehat{\mu}^{X\dagger}_{\bx+\ez}X_{\bx}\widehat{\tau}^{X}_{\bx-\ex}\widehat{\mu}^{X\dagger}_{\bx-\ey}\widehat{\mu}^{X}_{\bx-\ez}&=&1, \quad\forall\bx,\nonumber\\
            \widehat{\sigma}^{X\dagger}_{\by+\ey}\widehat{\nu}^{X\dagger}_{\by+\ex}\widehat{\nu}^{X\dagger}_{\by+\ez}X_{\by}\widehat{\sigma}^{X}_{\by-\ey}\widehat{\nu}^{X\dagger}_{\by-\ex}\widehat{\nu}^{X}_{\by-\ez}&=&1, \quad\forall\by,\nonumber\\
            \widehat{\eta}^{X\dagger}_{\bz+\ez}\widehat{\rho}^{X\dagger}_{\bz+\ex}\widehat{\rho}^{X\dagger}_{\bz+\ey}X_{\bz}\widehat{\eta}^{X}_{\bz-\ez}\widehat{\rho}^{X\dagger}_{\bz-\ex}\widehat{\rho}^{X}_{\bz-\ey}&=&1, \quad\forall\bz.\label{gauss6}
\end{eqnarray}
See also Fig.~\ref{3D3}. Accordingly, quadratic terms of spins in~\eqref{52} are modified as 
\begin{eqnarray}
    Z_{\bx+\exx}^{\dagger}   Z_{\bx}\to   Z_{\bx+\exx}^{\dagger}  \widehat{\tau}^Z_{\br} Z_{\bx},\quad Z_{\bx+\eyy}^{\dagger}   Z_{\bx}\to   Z_{\bx+\eyy}^{\dagger}  \widehat{\mu}^Z_{\bp_{xy}} Z_{\bx},\quad Z_{\bx+\ezz}^{\dagger}   Z_{\bx}\to   Z_{\bx+\ezz}^{\dagger}  \widehat{\mu}^Z_{\bp_{zx}} Z_{\bx}\nonumber\\
        Z_{\by+\eyy}^{\dagger}   Z_{\by}\to   Z_{\by+\eyy}^{\dagger}  \widehat{\sigma}^Z_{\br} Z_{\by},\quad Z_{\by+\exx}^{\dagger}   Z_{\by}\to   Z_{\by+\exx}^{\dagger}  \widehat{\nu}^Z_{\bp_{xy}} Z_{\by},\quad Z_{\by+\ezz}^{\dagger}   Z_{\by}\to   Z_{\by+\ezz}^{\dagger}  \widehat{\nu}^Z_{\bp_{zx}} Z_{\by}\nonumber\\
                Z_{\bz+\ezz}^{\dagger}   Z_{\bz}\to   Z_{\bz+\ezz}^{\dagger}  \widehat{\eta}^Z_{\br} Z_{\bz},\quad Z_{\bz+\exx}^{\dagger}   Z_{\bz}\to   Z_{\bz+\exx}^{\dagger}  \widehat{\rho}^Z_{\bp_{zx}} Z_{\bz},\quad Z_{\bz+\eyy}^{\dagger}   Z_{\bz}\to   Z_{\bz+\eyy}^{\dagger}  \widehat{\rho}^Z_{\bp_{yz}} Z_{\bz},\label{59}
\end{eqnarray}
in order for them to commute with Gauss's laws~\eqref{gauss6}.
\begin{figure}
    \hspace{15mm}
         \begin{subfigure}[h]{0.39\textwidth}
       \centering
  \includegraphics[width=2\textwidth]{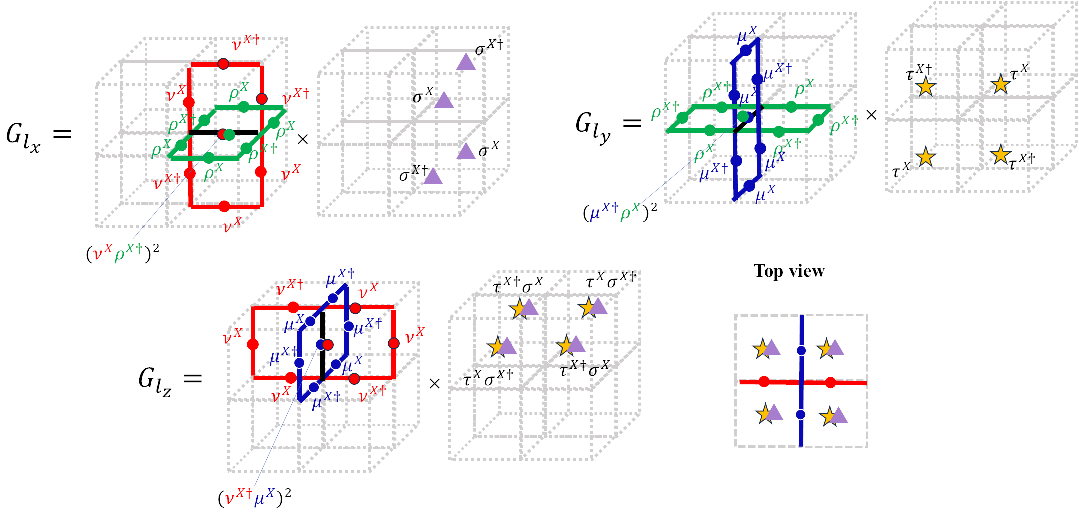}
         \caption{}\label{Glx}
             \end{subfigure}

             \hspace{15mm}
              \begin{subfigure}[h]{0.39\textwidth}
  \includegraphics[width=2\textwidth]{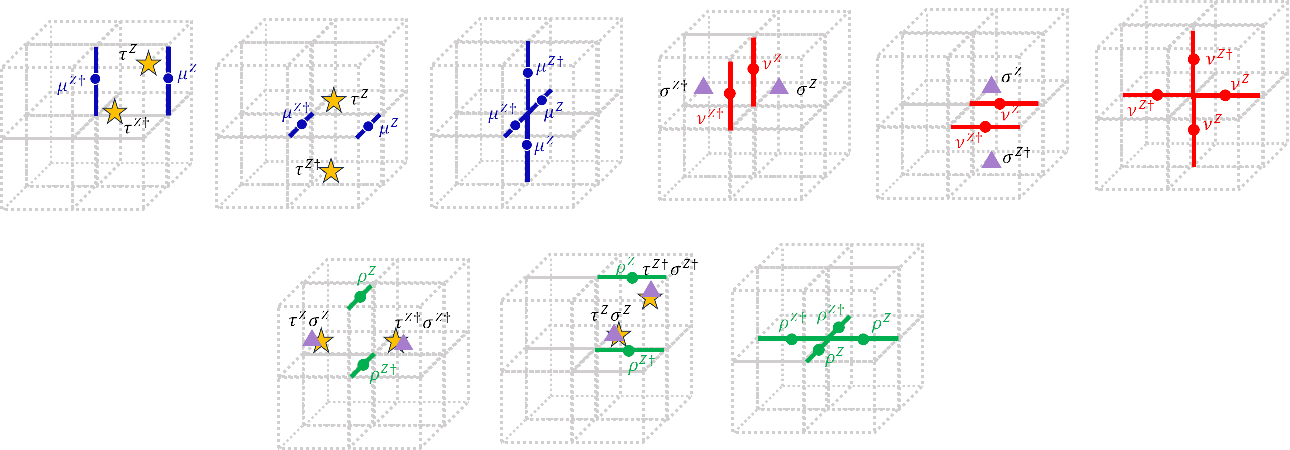}
         \caption{}\label{Glx2}
             \end{subfigure}  
             
             \hspace{35mm}
                     \begin{subfigure}[h]
                {0.39\textwidth}

  \includegraphics[width=0.95\textwidth]{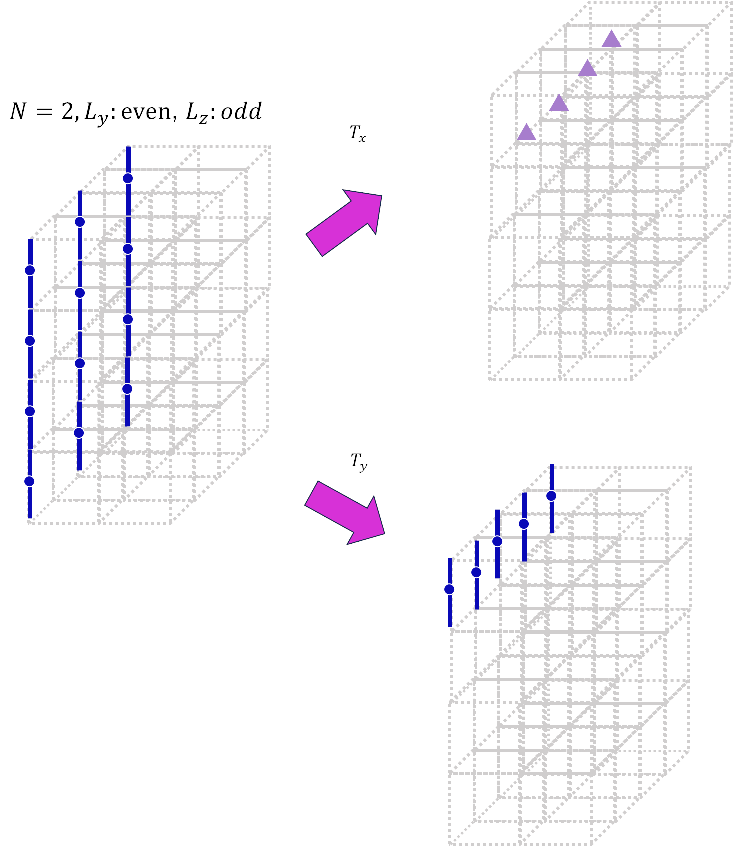}
         \caption{}\label{Glx3}
             \end{subfigure}

 \caption{(a)~Three terms in~\eqref{Gxyz}. The top view of $G_{\bz}$ from the $z$-axis is also shown.  
 (b)~Nine flux operators introduced in~\eqref{3dgauge}. (c)~An example of the dipole algebra in the case of $N=2$, even $L_y$ and odd $L_z$. When acting a translational operator on a noncontractible membrane operator, a loop operator is generated. 
 }
 \end{figure}
To proceed, we rewrite the operators as
\begin{eqnarray}
   &\tau^Z_{\br}\vcentcolon= Z_{\bx+\exx}^{\dagger}  \widehat{\tau}^Z_{\br} Z_{\bx},\quad
\mu_{\bp_{ab}}^Z\vcentcolon=Z_{\mathbf{l}_a+\mathbf{e}_{b}}^{\dagger}  \widehat{\mu}^Z_{\bp_{ab}} Z_{\mathbf{l}_a},\quad  \tau^X_{\br}\vcentcolon= \widehat{\tau}^X_{\br},\quad {\mu}^X_{\bp_{ab}}\vcentcolon=\widehat{\mu}^X_{\bp_{ab}}~[(a,b)=(y,x),(z,x)]\nonumber\\
  &\sigma^Z_{\br}\vcentcolon= Z_{\by+\eyy}^{\dagger}  \widehat{\sigma}^Z_{\br} Z_{\by},\quad
\nu_{\bp_{cd}}^Z\vcentcolon=Z_{\mathbf{l}_c+\mathbf{e}_{d}}^{\dagger}  \widehat{\nu}^Z_{\bp_{cd}} Z_{\mathbf{l}_c},\quad  \sigma^X_{\br}\vcentcolon= \widehat{\sigma}^X_{\br},\quad {\nu}^X_{\bp_{cd}}\vcentcolon=\widehat{\nu}^X_{\bp_{cd}}~[(c,d)=(x,y),(z,y)]\nonumber\\
  &\eta^Z_{\br}\vcentcolon= Z_{\bz+\ezz}^{\dagger}  \widehat{\eta}^Z_{\br} Z_{\bz},\quad
\rho_{\bp_{ef}}^Z\vcentcolon=Z_{\mathbf{l}_e+\mathbf{e}_{f}}^{\dagger}  \widehat{\rho}^Z_{\bp_{ef}} Z_{\mathbf{l}_e},\quad  \eta^X_{\br}\vcentcolon= \widehat{\eta}^X_{\br},\quad {\rho}^X_{\bp_{ef}}\vcentcolon=\widehat{\rho}^X_{\bp_{ef}}~[(e,f)=(x,z),(y,z)].\label{60}
\end{eqnarray}
We add the following gauge flux operators
\begin{eqnarray}
    -h_1\sum_{\by}\tau^Z_{\by+\ey}\mu^{Z\dagger}_{\by+\ex}\tau^{Z\dagger}_{\by-\ey}\mu^Z_{\by-\exx}-h_2\sum_{\bz}\tau^Z_{\bz+\ez}\mu^{Z\dagger}_{\bz+\ex}\tau^{Z\dagger}_{\bz-\ez}\mu^Z_{\bz-\ex}-h_3\sum_{\bc}\mu^{Z}_{\bc+\ey}\mu^{Z}_{\bc+\ez}\mu^{Z\dagger}_{\bc\ey}\mu^{Z\dagger}_{\bc\ez}+h.c.,
\end{eqnarray}
to the Hamiltonian~\eqref{52} to ensure that gauged theory is dynamically trivial. Further, referring to~\eqref{59} and~\eqref{60}, the fluxless condition~\eqref{fluxless} becomes
\begin{eqnarray}
    \eqref{fluxless}\leftrightarrow \tau^Z_{\br}\sigma^Z_{\br}\eta^Z_{\br}=1\quad\forall\br.
\end{eqnarray}
Substituting Gauss's laws~\eqref{gauss6} and~\eqref{60} into $\mathcal{P}^X_{ab}$ in~\eqref{52}, and changing the lattice grid so that we exchange $p$-cell and $(3-p)$-cell ($0\leq p\leq 3$) to make the model visually friendly, we finally arrive at the following gauged Hamiltonian:
\begin{eqnarray}
    \widehat{H}_{3D}&=&-J_p\left[\sum_{\bx}G_{\bx}+\sum_{\by}G_{\by}+\sum_{\by}G_{\by}\right]
    -J_x\left[\sum_{\by}\mu^{Z}_{\by}+\sum_{\bz}\mu^{Z}_{\bz}+\sum_{c}\tau^Z_{\bc}\right]
    -J_y\left[\sum_{\bx}\rho^{Z}_{\by}+\sum_{\bz}\rho^{Z}_{\bz}+\sum_{c}\sigma^Z_{\bc}\tau^Z_{\bc}\right]\nonumber\\
    &-&
    J_z\left[\sum_{\by}\nu^{Z}_{\bx}+\sum_{\bz}\nu^{Z}_{\bz}+\sum_{c}\sigma^Z_{\bc}\right]
    -J_{B_1}\sum_{\bp_{zx}}\prod_{s,t=\pm1}\left\{\left(\mu^{Z}_{\bp_{zx}+s\ex}\right)^s\left(\tau^{Z}_{\bp_{zx}+t\ey}\right)^t\right\}
    \nonumber\\
    &-&
J_{B_2}\sum_{\bp_{xy}}\prod_{s,t=\pm1}\left\{\left(\mu^{Z}_{\bp_{xy}+s\ex}\right)^s\left(\tau^{Z}_{\bp_{xy}+t\ez}\right)^t\right\}
-J_{B_3}\sum_{\br}\prod_{s,t=\pm1}\left\{\left(\mu^{Z}_{\br+s\ey}\right)^s\left(\mu^{Z}_{\br+t\ez}\right)^{-t}\right\}\nonumber\\
&-&J_{B_4}\sum_{\bp_{yz}}\prod_{s,t=\pm1}\left\{\left(\nu^{Z}_{\bp_{yz}+s\ex}\right)^s\left(\sigma^{Z}_{\bp_{yz}+t\ey}\right)^t\right\}
-J_{B_5}\sum_{\br}\prod_{s,t=\pm1}\left\{\left(\nu^{Z}_{\br+s\ex}\right)^s\left(\sigma^{Z}_{\br+t\ey}\right)^{-t}\right\}\nonumber\\
&-&J_{B_6}
\sum_{\br}\prod_{s,t=\pm1}\left\{\left(\mu^{Z}_{\br+s\ex}\right)^s\left(\mu^{Z}_{\br+t\ez}\right)^{-t}\right\}
-J_{B_7}\sum_{\bp_{yz}}\prod_{s,t=\pm1}\left\{\left(\rho^{Z}_{\bp_{yz}+s\ez}\right)^s\left(\tau^{Z}_{\bp_{yz}+t\ex}\sigma^{Z}_{\bp_{yz}+t\ex}\right)^{-t}\right\}\nonumber\\
&-&J_{B_8}\sum_{\bp_{zx}}\prod_{s,t=\pm1}\left\{\left(\rho^{Z}_{\bp_{zx}+s\ez}\right)^s\left(\tau^{Z}_{\br+t\ex}\sigma^{Z}_{\br+t\ex}\right)^{-t}\right\}\nonumber\\
&-&J_{B_9}
\sum_{\br}\prod_{s,t=\pm1}\left\{\left(\rho^{Z}_{\br+s\ex}\right)^s\left(\rho^{Z}_{\br+t\ey}\right)^{-t}\right\}+h.c.,\label{3dgauge}
\end{eqnarray}
where
\begin{eqnarray}
    G_{\bx}\vcentcolon&=&\nu^{X\dagger}_{\bx+\ezz}(\nu^{X}_{\bx})^2\nu^{X\dagger}_{\bx-\ezz}
   \times\rho^X_{\bx+\eyy}(\rho^{X\dagger}_{\bx})^2\rho^X_{\bx-\eyy}  \nonumber\\
    &&\times\prod_{s,t=\pm1 } \left(\nu^{X}_{\bx+s\ex+t\ez}
    \right)^{-st}\times
    \prod_{p,q=\pm1 } \left(\nu^{X}_{\bx+p\ex+q\ey}
    \right)^{pq}\times\prod_{a,b=\pm1}\left(\sigma^X_{\bx+a\ex+b\ey}\right)^{-ab}\nonumber\\
    G_{\by}\vcentcolon&=&\mu^{X}_{\by+\ezz}(\mu^{X\dagger}_{\by})^2\mu^{X\dagger}_{\by-\ezz}
   \times\rho^{X\dagger}_{\by+\exx}(\rho^{X}_{\by})^2\rho^{X\dagger}_{\by-\exx}  \nonumber\\
    &&\times\prod_{s,t=\pm1 } \left(\mu^{X}_{\by+s\ey+t\ez}
    \right)^{-st}\times
    \prod_{p,q=\pm1 } \left(\rho^{X}_{\by+p\ex+q\ey}
    \right)^{pq}\times\prod_{a,b=\pm1}\left(\tau^X_{\by+a\ex+b\ez}\right)^{ab}\nonumber\\
    G_{\bz}\vcentcolon&=&\mu^{X\dagger}_{\bz+\eyy}(\mu^{X}_{\bz})^2\mu^{X\dagger}_{\bz-\eyy}
   \times\nu^{X\dagger}_{\bz+\exx}(\nu^{X}_{\bz})^2\nu^{X\dagger}_{\bz-\exx}  \nonumber\\
    &&\times\prod_{s,t=\pm1 } \left(\mu^{X}_{\bz+s\ey+t\ez}
    \right)^{-st}\times
    \prod_{p,q=\pm1 } \left(\nu^{X}_{\bz+p\ez+q\ex}
    \right)^{pq}\times\prod_{a,b=\pm1}\left(\tau^X_{\bz+a\ex+b\ey}\sigma^{X\dagger}_{\bz+a\ex+b\ey}\right)^{ab}.\label{Gxyz}
\end{eqnarray}
The terms given in~\eqref{Gxyz} and nine flux operators defined in~\eqref{3dgauge} are portrayed in Fig.~\ref{Glx} and~\ref{Glx2}, respectively. Recall that we take the limit of $J_{B_i}\to\infty~(1\leq i\leq 9)$ so that the gauged theory does not admit fluxes.  
\par
We investigate what is the symmetry that the gauged Hamiltonian~\eqref{3dgauge} respects. We have the following $1$-form
modulated symmetries described by noncontractible membrane operators:
\begin{eqnarray}
    Q^{(1)xy,1}\vcentcolon=\left[\prod_{\hx=1}^{L_x}\prod_{\hy=1}^{L_y}\left(\rho^Z_{(\hx,\hy+\frac{1}{2},1)}\right)^{\hx}\right]^{\alpha_x},\quad Q^{(1)xy,2}\vcentcolon=\left[\prod_{\hx=1}^{L_x}\prod_{\hy=1}^{L_y}\left(\rho^Z_{(\hx+\frac{1}{2},\hy,1)}\right)^{\hy}\right]^{\alpha_y}\nonumber\\
    Q^{(1)yz,1}\vcentcolon=\left[\prod_{\hy=1}^{L_y}\prod_{\hz=1}^{L_z}\left(\mu^Z_{(1,\hy,\hz+\frac{1}{2})}\right)^{\hy}\right]^{\alpha_y},\quad Q^{(1)yz,2}\vcentcolon=\left[\prod_{\hy=1}^{L_y}\prod_{\hz=1}^{L_z}\left(\mu^Z_{(1,\hy+\frac{1}{2},\hz)}\right)^{\hz}\right]^{\alpha_z}\nonumber\\
     Q^{(1)zx,1}\vcentcolon=\left[\prod_{\hz=1}^{L_z}\prod_{\hx=1}^{L_x}\left(\nu^Z_{(\hx+\frac{1}{2},1,\hz)}\right)^{\hz}\right]^{\alpha_z},\quad Q^{(1)zx,2}\vcentcolon=\left[\prod_{\hz=1}^{L_z}\prod_{\hx=1}^{L_x}\left(\nu^Z_{(\hx,1,\hz+\frac{1}{2})}\right)^{\hx}\right]^{\alpha_x}.\label{ffff}
\end{eqnarray}
Here, we have defined $\alpha_{i}=\frac{N}{\gcd(N,L_i)}~(i=x,y,z)$. 
Depending on $N$, and the system size, the first and second charges in~\eqref{ffff} are not independent: 
If $\gcd(N,L_x)$ and $\gcd(N,L_y)$ are more than one, further, there exit integers $\{c_i:1\leq c_i\leq \gcd(N,L_i)-1,~i=x,y\}$ such that $c_x\alpha_x+c_y\alpha_y=0\mod N$, then 
    by the fluxless condition, it follows that 
    the two charges are subject to
$\left[Q^{(1)xy,1}\right]^{c_x}\times \left[Q^{(1)xy,2}\right]^{c_y} =I$.
Likewise, regarding the third and fourth charges, if $\gcd(N,L_y)$ and $\gcd(N,L_z)$ are more than one, and there exit integers $\{c_i:1\leq c_i\leq \gcd(N,L_i)-1,~i=y,z\}$ such that $c_y\alpha_y+c_z\alpha_z=0\mod N$, then 
    by the fluxless condition, we have
$\left[Q^{(1)yz,1}\right]^{c_y}\times \left[Q^{(1)yz,2}\right]^{c_z} =I$. Wee have the similar relation for the last two charges in~\eqref{ffff}:
If  $\gcd(N,L_z)$ and $\gcd(N,L_x)$ are more than one, and there exit integers $\{c_i:1\leq c_i\leq \gcd(N,L_i)-1,~i=z,x\}$ so that $c_z\alpha_z+c_x\alpha_x=0\mod N$, then we have $\left[Q^{(1)zx,1}\right]^{c_z}\times \left[Q^{(1)zx,2}\right]^{c_x} =I$.\par

The model~\eqref{3dgauge} also admits $2$-form symmetries corresponding to the following noncontractible loops:
\begin{eqnarray}
    Q^{(2)x,1}\vcentcolon=\prod_{\hx=1}^{L_x}\tau^Z_{(\hx+\frac{1}{2},\frac{1}{2},\frac{1}{2})},\quad Q^{(2)x,2}\vcentcolon=\prod_{\hx=1}^{L_x}\nu^Z_{(\hx,1,\frac{1}{2})},\quad Q^{(2)x,3}\vcentcolon=\prod_{\hx=1}^{L_x}\rho^Z_{(\hx,\frac{1}{2},1)}\nonumber\\
Q^{(2)y,1}\vcentcolon=\prod_{\hy=1}^{L_y}\sigma^Z_{(\frac{1}{2},\hy+\frac{1}{2},\frac{1}{2})},\quad Q^{(2)y,2}\vcentcolon=\prod_{\hy=1}^{L_y}\rho^Z_{(\frac{1}{2},\hy,1)},\quad Q^{(2)y,3}\vcentcolon=\prod_{\hy=1}^{L_y}\mu^Z_{(1,\hy,\frac{1}{2})},\nonumber\\
Q^{(2)z,1}\vcentcolon=\prod_{\hz=1}^{L_z}\tau^Z_{(\frac{1}{2},\frac{1}{2},\hz+\frac{1}{2})}\sigma^Z_{(\frac{1}{2},\frac{1}{2},\hz+\frac{1}{2})},\quad Q^{(2)z,2}\vcentcolon=\prod_{\hz=1}^{L_z}\mu^Z_{(1,\hy+\frac{1}{2},\hz)},\quad Q^{(2)z,3}\vcentcolon=\prod_{\hz=1}^{L_z}\nu^Z_{(\frac{1}{2},1,\hz)}.
\end{eqnarray}
The $1$-form and $2$-form symmetries are related via translational operators. To wit, 
\begin{eqnarray}
T_y Q^{(1)xy,2}T_y^{-1}=Q^{(1)xy,2} \left(Q^{(2)y,2\dagger}\right)^{\alpha_yL_x},\quad  T_z Q^{(1)xy,2}T_z^{-1}=Q^{(1)xy,2} \left(Q^{(2)y,1\dagger}\right)^{\alpha_yL_x},\nonumber\\
   T_z Q^{(1)zx,1}T_z^{-1}=Q^{(1)zx,1} \left(Q^{(2)z,3\dagger}\right)^{\alpha_zL_x},\quad  T_y Q^{(1)zx,1}T_y^{-1}=Q^{(1)zx,1} \left(Q^{(2)z,1}\right)^{\alpha_zL_x}\nonumber\\
T_z Q^{(1)yz,2}T_z^{-1}=Q^{(1)yz,2} \left(Q^{(2)z,2\dagger}\right)^{\alpha_zL_y},\quad  T_x Q^{(1)yz,2}T_x^{-1}=Q^{(1)yz,2} \left(Q^{(2)z,1}\right)^{\alpha_zL_y},\nonumber\\
      T_x Q^{(1)xy,1}T_x^{-1}=Q^{(1)xy,1} \left(Q^{(2)x,3\dagger}\right)^{\alpha_xL_y},\quad  T_z Q^{(1)xy,1}T_z^{-1}=Q^{(1)xy,1} \left(Q^{(2)x,1\dagger}\right)^{\alpha_xL_y},\nonumber\\
      T_x Q^{(1)zx,2}T_x^{-1}=Q^{(1)zx,2} \left(Q^{(2)x,2\dagger}\right)^{\alpha_xL_z},\quad  T_yQ^{(1)zx,2}T_y^{-1}=Q^{(1)zx,2} \left(Q^{(2)x,1\dagger}\right)^{\alpha_xL_z},\nonumber\\
          T_y Q^{(1)yz,1}T_y^{-1}=Q^{(1)yz,1} \left(Q^{(2)y,3\dagger}\right)^{\alpha_yL_z},\quad  T_x Q^{(1)yz,1}T_x^{-1}=Q^{(1)yz,1} \left(Q^{(2)y,1\dagger}\right)^{\alpha_yL_z}.\label{68}
\end{eqnarray}
Acting a translational operator on a $1$-form symmetry yields stack of~$2$-form symmetries,
manifested as the power $L_I$ on the right hand side of relations in~\eqref{68}. We demonstrate one of the relations
in Fig.~\ref{Glx3} with $N=2$, even $L_y$ and odd $L_z$. We obtain a new dipole algebra consisting of $1$-form and $2$-form symmetries.

\subsection{Gauging $1$-form symmetry}
\begin{figure}
\hspace{40mm}
         \begin{subfigure}[h]{0.35\textwidth}
       \centering
  \includegraphics[width=1.4\textwidth]{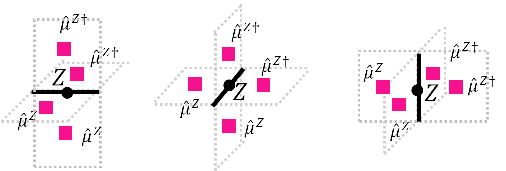}
         \caption{}\label{tr1}
             \end{subfigure}
            
            \hspace{30mm}\begin{subfigure}[h]{0.55\textwidth}
            \centering
  \includegraphics[width=0.9\textwidth]{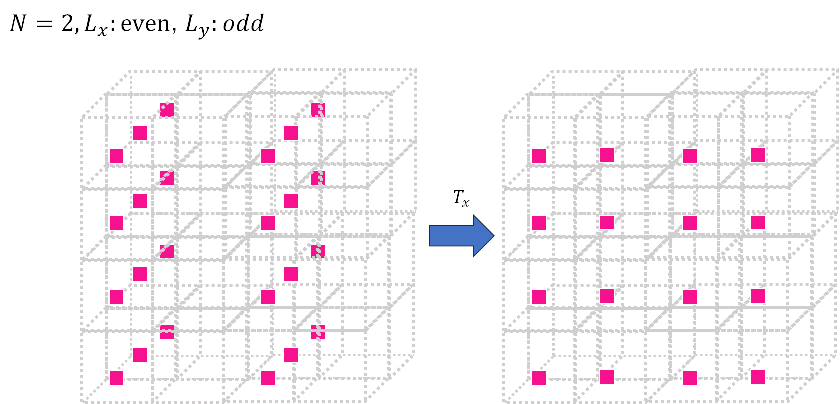}
         \caption{}\label{tr2}
             \end{subfigure}

 \caption{(a)~Gauss's laws corresponding to~\eqref{gauss_final}.
 (b) The action of translational operator in the $x$-direction on $0$-form symmetry~$Q^{(0)zx,2}$ (left) gives rise to $1$-form symmetry, $Q^{(1)zx}$~(right)~in the case of $N=2$, even $L_x$ and odd $L_y$. 
 }
 \end{figure}

In this subsection, we turn to gauging $1$-form symmetry~\eqref{membrane} in the model~\eqref{52}. 
To this end, 
we introduce extended Hilbert space on each plaquette whose $\mathbb{Z}_{N}$ Pauli operator is denoted as $\widehat{\mu}_{\bp_{ab}}^{X/Z}$. Gauss's laws are given by
\begin{eqnarray}
    Z_{\mathbf{l}_a}\prod_{s,t=\pm1}\left(\widehat{\mu}_{\mathbf{l}_a+s\mathbf{e}_b}^{Z\dagger}\right)^{s}\left(\widehat{\mu}_{\mathbf{l}_a+t\mathbf{e}_c}^{Z\dagger}\right)^{t}=1,\label{gauss_final}
\end{eqnarray}
where $(a,b,c)$ are cyclic permutation of $(x,y,z)$ (See also Fig.~\ref{tr1}). We modify the term $\mathcal{P}_{\bp_{ab}}$ defined in~\eqref{3dtoric} so that it commutes with Gauss's laws:
\begin{eqnarray}
    \mathcal{P}_{\bp_{ab}}\to \mathcal{P}_{\bp_{ab}}\widehat{\mu}^X_{\bp_{ab}}.
\end{eqnarray}
To proceed, we rewrite the operators as
\begin{eqnarray}
\mu^Z_{\bp_{ab}}\vcentcolon=\widehat{\mu}^Z_{\bp_{ab}},\quad \mu^X_{\bp_{ab}}\vcentcolon= \mathcal{P}_{\bp_{ab}}\widehat{\mu}^X_{\bp_{ab}}.
\end{eqnarray}
Further, we add the following flux operator
\begin{eqnarray}
  -g_{f}\sum_{\bc}\prod_{s=\pm1}\prod_{t\pm1}\prod_{u\pm1}\left( \mu^X_{\bc+s\exx}\right)^s\left(\mu^{X\dagger}_{\bc+t\eyy} \right)^t\left(\mu^X_{\bc+u\ezz}\right)^u+h.c.
\end{eqnarray}
to the Hamiltonian~\eqref{52}
 so that theory is dynamically trivial. Overall, the gauged Hamiltonian reads
 \begin{eqnarray}
     \tilde{H}_{3D}=
&-&\sum_{a=x,y,z}J_a\left[\sum_{\mathbf{l}_a}\left(G^{Z\dagger}_{\mathbf{l}_a+\exx}G^Z_{\mathbf{l}_a}+G^{Z\dagger}_{\mathbf{l}_a+\eyy}G^Z_{\mathbf{l}_a}+G^{Z\dagger}_{\mathbf{l}_a+\ezz}G^Z_{\mathbf{l}_a}\right)\right]
\nonumber\\
&-&g_{f}\sum_{\bc}\prod_{s=\pm1}\prod_{t\pm1}\prod_{u\pm1}\left( \mu^X_{\bc+s\exx}\right)^s\left(\mu^{X\dagger}_{\bc+t\eyy} \right)^t\left(\mu^X_{\bc+u\ezz}\right)^u+h.c.,\label{final}
 \end{eqnarray}
where 
 \begin{eqnarray}
     G^Z_{\mathbf{l}_a}\vcentcolon=\prod_{s,t=\pm1}\left({\mu}_{\mathbf{l}_a+s\mathbf{e}_b}^{Z}\right)^{s}\left({\mu}_{\mathbf{l}_a+t\mathbf{e}_c}^{Z}\right)^{t}.
 \end{eqnarray}
 While the terms in the second line of~\eqref{final} describe flux operators that were introduced in the 3D toric code, the ones in the first line correspond to the product of the adjacent ``star operators",~$G_{\mathbf{l}_a}^Z$. 
 In the following, we take $g_f\to\infty$ so that the model does not admit any flux.
 \par
 The model~\eqref{final} respects the following $0$-form modulated symmetries generated by
 \begin{eqnarray}
     Q^{(0)xy,1}\vcentcolon=\left[\prod_{\hx=1}^{L_x}\prod_{\hy=1}^{L_y}\prod_{\hz=1}^{L_z} (\mu^X_{\bp_{xy}})^{\hx}\right]^{\alpha_x},\quad Q^{(0)xy,2}\vcentcolon=\left[\prod_{\hx=1}^{L_x}\prod_{\hy=1}^{L_y}\prod_{\hz=1}^{L_z} (\mu^X_{\bp_{xy}})^{\hy}\right]^{\alpha_y}\nonumber\\
     Q^{(0)yz,1}\vcentcolon=\left[\prod_{\hx=1}^{L_x}\prod_{\hy=1}^{L_y}\prod_{\hz=1}^{L_z} (\mu^X_{\bp_{yz}})^{\hy}\right]^{\alpha_y},\quad Q^{(0)yz,2}\vcentcolon=\left[\prod_{\hx=1}^{L_x}\prod_{\hy=1}^{L_y}\prod_{\hz=1}^{L_z} (\mu^X_{\bp_{yz}})^{\hz}\right]^{\alpha_z}\nonumber\\
     Q^{(0)zx,1}\vcentcolon=\left[\prod_{\hx=1}^{L_x}\prod_{\hy=1}^{L_y}\prod_{\hz=1}^{L_z} (\mu^X_{\bp_{zx}})^{\hz}\right]^{\alpha_z},\quad Q^{(0)zx,2}\vcentcolon=\left[\prod_{\hx=1}^{L_x}\prod_{\hy=1}^{L_y}\prod_{\hz=1}^{L_z} (\mu^X_{\bp_{zx}})^{\hx}\right]^{\alpha_x}.\label{0_final}
 \end{eqnarray}
Depending on $N$ and system size, there are several constraints on these charges: If $\gcd(N,L_x),\gcd(N,L_y)>1$, and there exit integers $\{c_i:1\leq c_i\leq \gcd(N,L_i)-1,~i=x,y\}$ such that $c_x\alpha_x-c_y\alpha_y=0\mod N$, then we have $\left[Q^{(0)yz,1}\right]^{c_y}=\left[Q^{(0)zx,2}\right]^{c_x} $. Also, if $\gcd(N,L_y),\gcd(N,L_z)>1$, and there exit integers $\{c_i:1\leq c_i\leq \gcd(N,L_i)-1,~i=y,z\}$ such that $c_y\alpha_y+c_z\alpha_z=0\mod N$, then we have $\left[Q^{(0)xy,2}\right]^{c_y}\times\left[Q^{(0)zx,1}\right]^{c_z}=I $. The similar constraint $\left[Q^{(0)xy,1}\right]^{c_x}\times\left[Q^{(0)yz,1}\right]^{c_z}=I $ can be obtained if $\gcd(N,L_z),\gcd(N,L_x)>1$, and there exit integers $\{c_i:1\leq c_i\leq \gcd(N,L_i)-1,~i=z,x\}$ so that $c_z\alpha_z+c_x\alpha_x=0\mod N$.\par

The model also respects the following $1$-form symmetries generated by the noncontractible membranes:
\begin{eqnarray}
    Q^{(1)xy}\vcentcolon=\prod_{\hx=1}^{L_x}\prod_{\hy=1}^{L_y}\mu^X_{(\hx+\frac{1}{2},\hy+\frac{1}{2},1)},\quad     Q^{(1)yz}\vcentcolon=\prod_{\hy=1}^{L_y}\prod_{\hz=1}^{L_z}\mu^X_{(1,\hy+\frac{1}{2},\hz+\frac{1}{2})},\quad Q^{(1)zx}\vcentcolon=\prod_{\hz=1}^{L_z}\prod_{\hx=1}^{L_x}\mu^X_{(\hx+\frac{1}{2},1,\hz+\frac{1}{2})}.\label{1_final}
\end{eqnarray}
Furthermore, the $0$-form and $1$-form symmetry operators are related via translational operators
\begin{eqnarray}
T_xQ^{(0)xy,1}T_x^{-1}=Q^{(0)xy,1}\left(Q^{(1)xy\dagger}\right)^{\alpha_xL_z},\quad T_yQ^{(0)xy,2}T_y^{-1}=Q^{(0)xy,2}\left(Q^{(1)xy\dagger}\right)^{\alpha_yL_z}\nonumber\\
T_yQ^{(0)yz,1}T_y^{-1}=Q^{(0)yz,1}\left(Q^{(1)yz\dagger}\right)^{\alpha_yL_x},\quad T_zQ^{(0)yz,2}T_z^{-1}=Q^{(0)yz,2}\left(Q^{(1)yz\dagger}\right)^{\alpha_zL_x}\nonumber\\
T_zQ^{(0)zx,1}T_z^{-1}=Q^{(0)zx,1}\left(Q^{(1)zx\dagger}\right)^{\alpha_zL_y},\quad T_xQ^{(0)zx,2}T_x^{-1}=Q^{(0)zx,2}\left(Q^{(1)zx\dagger}\right)^{\alpha_xL_y}\label{76}
\end{eqnarray}
We demonstrate the last relation in~\eqref{76} with $N=2$, even $L_x$ and odd $L_y$ in Fig.~\ref{tr2}. We obtain a new and unusual dipole algebra consisting of $0$-form and $1$-form symmetries.

\section{Gauge fields for 0-form U(1) global symmetry}\label{u1}
In this section, we review how to gauge standard global U(1) symmetry in view of a field theory, 
which may help one to understand gauging dipole symmetry discussed in Sec.~\ref{sec:51}. 
To start, we consider
a theory with global U(1) zero-form symmetry (i.e., symmetry operation acts on an entire space) and its charge $Q$ defined in $(d+1)$ spatial dimension.
The conserved charge is expressed as 
\begin{equation}
    Q(V)=\int_{V_{d}} * j, 
\end{equation}
where $j$ and $V$ denotes the conserved 1-form current and $d$ dimensional spatial volume, and $*$ does the Hodge dual.
This charge commutes with the translation operation,~$P_{I}\;(I=1,\cdots,d)$, i.e., 
\begin{equation}
   \left[ iP_I,Q \right] =0
 \label{eq:re}.
\end{equation}
We introduce a one-form U(1) gauge field $a$ which couples with the current $j$ with the coupling term being described by 
\begin{eqnarray}
    S_{c}=\int_Va\wedge* j\label{cp}.
\end{eqnarray}
With the gauge transformation ($\chi$: gauge parameter)
   $ a\ \to\  a+d\chi$
and the condition that the coupling term~\eqref{cp} is gauge invariant, we have the conservation law of the current $d* j=0$.\par
In the main text, we apply this logic to the case of dipole symmetry, meaning, we define charges associated with dipole symmetry and express them in terms of the currents. Introducing gauge fields and coupling terms, we demand gauge transformation for the gauge fields so that gauge invariance of the coupling term leads to the conservation of the current. 
\section{Higher group from gauging with a mixed \tht~anomaly}\label{app2}
In this section, we review how the higher group is generalized. 
Given finite groups, $N$, $K$, and $G$, we start with the 
following exact sequence
\begin{eqnarray}
    1\to N\to G\to K\to 1,
\end{eqnarray}
whose central extension is characterized by $\epsilon\in H^2(K,N)$ with $H^p(K,N)$ being $p$-th cohomology group of $K$ with coefficients in $N$.
We introduce a theory in $(d+1)$-spacetime dimension where there are $(d-1)$-form $N$
and $0$-form $K$ symmetries with corresponding gauge fields being ${\alpha}^{(d)}$ and $\beta^{(1)}$.
Here,~$\alpha^{(d)}\in C^{d}(M,N),\beta^{(1)}\in C^{1}(M,K)$ [$C^p(M,N)$ represents $p$-th cochain of manifold $M$ with coefficients in $N$ and similarly for $C^p(M,K)$].
We further assume that $N$ is Abelian and that
the two global symmetries, $N$ and $K$ have the following~\tht~anomaly:
\begin{eqnarray}
    S=\int_{M_{d+2}}{\alpha}^{(d)}\cup \epsilon (\beta^{(1)}).\label{anomaly}
\end{eqnarray}
It is known that after gauging $p$-form symmetry in $(d+1)$-spacetime dimensions, 
there is a dual $(d-p-1)$-form symmetry~\cite{gaiotto2015generalized}. 
By gauging the $(d-1)$-form symmetry, the gauged partition function reads
\begin{eqnarray}
    Z[\widehat{\alpha}^{(1)},\beta^{(1)}]\sim\sum_{\alpha^{(d)}}Z[{\alpha}^{(d)},\beta^{(1)}]\exp\left[i\int_{M_{d+2}}{\alpha}^{(d)}\cup \epsilon (\beta^{(1)})\right]\times\exp\left[i\int_{M_{d+1}}{\alpha}^{(d)}\cup \widehat{\alpha}^{(1)}\right]\label{aA}
\end{eqnarray}
Here, $\widehat{\alpha}^{(1)}\in C^{1}(M,K)$ denotes the background gauge field corresponding to the dual $0$-form symmetry.
To make the theory~\eqref{aA} gauge invariant, we demand that 
\begin{eqnarray}
       d\widehat{\alpha}^{(1)}=\epsilon (\beta^{(1)}),\quad d\beta^{(1)}=0.\label{78}
\end{eqnarray}
While we have usual flatness condition of the gauge field~$\beta^{(1)}$, there is an unconventional flatness condition of the gauge field~$\widehat{\alpha}^{(1)}$, implying the nontrivial central extension.
\par
To recap the argument, if we start with a theory
with two global symmetries, $(d-1)$-form ${N}$
and $0$-form $K$ symmetries with \tht~anomaly determined by $\epsilon$~\eqref{anomaly}, and gauge one of the symmetries, $(d-1)$-form ${N}$ symmetry, we obtain a condition~\eqref{78}, where one of the flatness condition of a  gauge field is modified. 
In the main text, we introduce gauge fields of dipole symmetries~(Sec.~\ref{sec:51}). There is a resemblance between the condition~\eqref{78} and flatness condition of the gauge fields of dipole symmetries, implying that the emergence of the dipole symmetries can be accounted by the \tht~anomaly counter term in the similar form as~\eqref{anomaly}.

\section{Field theoretical analysis for 3D example }\label{3dth}
In this section, we provide field theoretical analysis on 3D lattice model studied in App.~\ref{sec4}. 
For one $1$-form and three $0$-form symmetries with conserved currents $*j\fl2$ and $*K\fii1~(I=1,2,3)$, their charges are defined as
\begin{eqnarray}
Q[\Sigma_3,\Sigma_2,e^x]=\int_{\Sigma_3}\left(*j\fl2\right)_{\Sigma_2}\wedge e^x,\quad Q^I[\Sigma_3]=\int_{\Sigma_3}*K\fii{1}~(I=1,2,3).\label{charges4}
\end{eqnarray}
Similar to the case of 2D, the first charge is defined in such a way that the current $*j\fl2$ defined on $\Sigma_2$ is stacked along the $x$-direction. 
We retain such dependence in the following argument, meaning, 
we write the two types of charges in~\eqref{charges4} as $Q[\Sigma_2,e^x]$ and $Q^I$.
We assume that 
\begin{eqnarray}
[\mathrm{i}P_x,Q^I]=0~(I=1,2,3),\quad [\mathrm{i}P_I,Q^J]=\delta_{I,J}Q[\Sigma_2,e^x]~(I,J=2,3),
[iP_I,Q[\Sigma_2,e^x]]=0~(I=1,2,3).\label{23}
\end{eqnarray}
To proceed, we can rewrite the current $*K\fii1$
as 
\begin{eqnarray}
    *K^{1(1)}&=&*k^{1(1)}-y\left(*j\fl2\right)_{\Sigma_2}\wedge e^z\\
     *K^{2(1)}&=&*k^{2(1)}-z\left(*j\fl2\right)_{\Sigma_2}\wedge e^x\\
      *K^{3(1)}&=&*k^{3(1)}-y\left(*j\fl2\right)_{\Sigma_2}\wedge e^x
\end{eqnarray}
to reproduce the relations~\eqref{23}. Here, $*k\fii1~(I=1,2,3)$ denotes nonconserved current. Introducing gauge fields $a\fl2$ and $A\fii1$, we define coupling term as
\begin{eqnarray}
    \Tilde{S}=\int_{V_4}\left(a\fl2\wedge *j\fl2+\sum_{I=1}^3A\fii2\wedge*k^{I(1)} \right).\label{cp3}
\end{eqnarray}
The following gauge transformation
\begin{eqnarray}
    a\fl2&\to& a\fl2+d\lambda\fl1+\Lambda^{1(0)}e^x\wedge e^y+\Lambda^{2(0)}e^y\wedge e^z+\Lambda^{3(0)}e^z\wedge e^x,\nonumber\\
    A\fii1&\to& A\fii1+d\Lambda\fii0~(I=1,2,3),
\end{eqnarray}
jointly with the gauge invariance of~\eqref{cp3} leads to the conservation law of the currents, that is, $d*j\fl2=d*K\fii1=0$. We introduce gauge invariant fluxes as 
\begin{eqnarray*}
    \tilde{f}_a\fl3&=&da\fl2-A^{1(1)}\wedge e^x\wedge e^y-A^{2(1)}\wedge e^y\wedge e^z-A^{3(1)}\wedge e^z\wedge e^x\nonumber\\
    \tilde{F}\fii2_A&=&dA\fii1,
\end{eqnarray*}
from which the flatness condition of the gauge fields is given by
\begin{eqnarray}
    da\fl2=A^{1(1)}\wedge e^x\wedge e^y+A^{2(1)}\wedge e^y\wedge e^z+A^{3(1)}\wedge e^z\wedge e^x,\quad dA\fii1=0~(I=1,2,3).\label{aaa}
\end{eqnarray}
Instead of~\eqref{charges4}, had we defined charges as (the foliation field $e^x$ is replaced with $e^y$)
\begin{eqnarray}
Q[\Sigma_3,\Sigma_2,e^y]=\int_{\Sigma_3}\left(*j\fl2\right)_{\Sigma_2}\wedge e^y,\quad Q^I[\Sigma_3]=\int_{\Sigma_3}*K\fii{1}~(I=1,2,3),\label{chages2} 
\end{eqnarray}
with relation 
\begin{eqnarray}
[iP_y,Q^1]=0~(I=1,2,3),\quad [iP_I,Q^J]=\delta_{I,J}Q[\Sigma_2,e^x]~(I,J=1,3),\quad
[iP_I,Q[\Sigma_2,e^y]]=0~(I=1,2,3),\label{24}
\end{eqnarray}
and introduced gauge fields associated with the symmetries, 
we would arrive at the same flatness condition of the gauge fields as~\eqref{aaa}. 
Likewise, if we define [we replace $e^x$ with $e^z$ compared with~\eqref{charges4}]
\begin{eqnarray}
Q[\Sigma_3,\Sigma_2,e^z]=\int_{\Sigma_3}\left(*j\fl2\right)_{\Sigma_2}\wedge e^z,\quad Q^I[\Sigma_3]=\int_{\Sigma_3}*K\fii{1}~(I=1,2,3),\label{chages3} 
\end{eqnarray}
with 
\begin{eqnarray}
[iP_z,Q^1]=0~(I=1,2,3),\quad [iP_I,Q^J]=\delta_{I,J}Q[\Sigma_2,e^z]~(I,J=1,2),\quad
[iP_I,Q[\Sigma_2,e^z]]=0~(I=1,2,3),\label{25}
\end{eqnarray}
and introduce gauge fields, 
we would end up with the same condition as~\eqref{aaa}. \par
We also discuss another type of dipole algebra in 3D, involving one $1$-form and three $2$-form symmetries. 
We consider a theory with such symmetries whose conserved currents are represented by $\widehat{K}\fl2$ and $\widehat{j}\fii3~(I=1,2,3)$. We introduce
\begin{eqnarray}
    \widehat{Q}_I[\Sigma_2,\Sigma_1,e^x]=\int_{\Sigma_3}\left(*\widehat{j}\fii3\right)_{\Sigma_1}\wedge e^x~(I=1,2,3),\quad \widehat{Q}[\Sigma_2]=\int_{\Sigma_2}*\widehat{K}\fl2 \label{final118}
\end{eqnarray}
with the following relation
\begin{eqnarray}
    [iP_I,\widehat{Q}_J[\Sigma_1,e^x]]=0~(I,J=1,2,3),\quad [iP_x,\widehat{Q}]=0,\quad [iP_I,\widehat{Q}]=\widehat{Q}_I[\Sigma_1,e^x]~(I=2,3).\label{iii}
\end{eqnarray}
One can rewrite the current $  *\widehat{K}\fl2$ as
\begin{eqnarray}
  *\widehat{K}\fl2=*\widehat{k}\fl2-y\left(*\widehat{j}^{1(3)}\right)_{\Sigma_1}\wedge e^{z}-y\left(*\widehat{j}^{2(3)}\right)_{\Sigma_1}\wedge e^x -z*\left(*\widehat{j}^{3(3)}\right)_{\Sigma_1}\wedge e^x 
\end{eqnarray}
which reproduces the relation~\eqref{iii}. Introducing gauge fields as $b\fii3$ and $B\fl2$, we define a coupling term as
\begin{eqnarray}
    S^\prime=\int_{V_4}\left( \sum_{I=1}^3b\fii3\wedge *\widehat{j}\fii3+B\fl2\wedge *\widehat{k}\fl2\right).\label{cpay}
\end{eqnarray}
The following gauge transformation 
\begin{eqnarray}
    b^{1(3)}&\to& b^{1(3)}+d\lambda^{1(2)}+
    \Lambda^{(1)}\wedge e^x\wedge e^y,\nonumber\\
    b^{2(3)}&\to& b^{2(3)}+d\lambda^{2(2)}+
    \Lambda^{(1)}\wedge e^y\wedge e^z,\nonumber\\
    b^{3(3)}&\to& b^{3(3)}+d\lambda^{3(2)}+
    \Lambda^{(1)}\wedge e^z\wedge e^x,\nonumber\\
    B\fl2&\to& B\fl2 +d\Lambda\fl1,
\end{eqnarray}
together with the gauge invariance of the coupling term~\eqref{cpay} yields the conservation law of the currents, viz, $d*\widehat{j}\fii3=d*K\fl2=0$.
Analogous to the previous arguments, one could introduce gauge invariant fluxes, from which the flatness condition of the gauge fields are given by
\begin{eqnarray}
      db^{1(3)}&=& 
    B\fl2\wedge e^x\wedge e^y,\nonumber\\
     db^{2(3)}&=& 
    B\fl2\wedge e^y\wedge e^z,\nonumber\\
     db^{3(3)}&=& 
    B\fl2\wedge e^z\wedge e^x,\nonumber\\
    dB\fl2&=&0.\label{bB}
\end{eqnarray}
\par
We could introduce other charges than the ones in~\eqref{final118} by replacing the foliation field $e^x$ with $e^y$ or~$e^z$. For instance, instead of~\eqref{final118}, if we introduce 
\begin{eqnarray}
    \widehat{Q}_I[\Sigma_2,\Sigma_1,e^y]=\int_{\Sigma_3}\left(*\widehat{j}\fii3\right)_{\Sigma_1}\wedge e^y~(I=1,2,3),\quad \widehat{Q}[\Sigma_2]=\int_{\Sigma_2}*\widehat{K}\fl2 \label{final119}
\end{eqnarray}
with relation
\begin{eqnarray}
    [iP_I,\widehat{Q}_J[\Sigma_1,e^y]]=0~(I,J=1,2,3),\quad [iP_y,\widehat{Q}]=0,\quad [iP_I,\widehat{Q}]=\widehat{Q}_I[\Sigma_1,e^y]~(I=1,3),\label{iiii}
\end{eqnarray}
the similar line of thoughts leads to that we have the same flatness condition of the gauge fields~\eqref{bB} when gauging dipole symmetry. Likewise, had we defined 
\begin{eqnarray}
    \widehat{Q}_I[\Sigma_2,\Sigma_1,e^z]=\int_{\Sigma_3}\left(*\widehat{j}\fii3\right)_{\Sigma_1}\wedge e^z~(I=1,2,3),\quad \widehat{Q}[\Sigma_2]=\int_{\Sigma_2}*\widehat{K}\fl2 
\end{eqnarray}
with relation
\begin{eqnarray}
    [iP_I,\widehat{Q}_J[\Sigma_1,e^z]]=0~(I,J=1,2,3),\quad [iP_z,\widehat{Q}]=0,\quad [iP_I,\widehat{Q}]=\widehat{Q}_I[\Sigma_1,e^z]~(I=1,2),\label{iiiii}
\end{eqnarray}
then we would arrive  at the same condition as~\eqref{bB} when gauging dipole symmetry.
\par
Now we are in a good place to study the relation between dipole symmetries that we have discussed in this subsection and the anomaly inflow counter term. 
We consider a theory in $(3+1)d$ with three $0$-form and one $1$-form $\mathbb Z_N$ symmetries whose corresponding gauge fields are represented by $G\fii1~(I=1,2,3)$, $H\fl2$, respectively. We assume these symmetries have mixed~\tht~anomaly, described by
\begin{eqnarray}
\boxed{S=\frac{iN}{2\pi}\int_{M_5}\left(\sum_{\substack{I,J,K=1,2.3 \\ I\neq J\neq K}}G\fii1\wedge H\fl2\wedge e^J\wedge e^K\right).\label{IJK}}
\end{eqnarray}
Here, the indices $I,J,K$ are cyclic permutation of $1,2,3$. The term~\eqref{IJK} indicates the mixed anomaly between $0$-form and $1$-form global symmetries, and translational symmetries in the $J$- and $K$-direction. 
\par
Gauging $0$-form symmetries, gives rise to dual~$2$-form symmetries with corresponding gauge fields being $\tilde{G}\fii3~(I=1,2,3)$. By following the similar argument presented in the previous subsections, we have the following flatness condition of the gauge fields:
\begin{eqnarray}
     d\tilde{G}^{1(3)}&=& 
    H\fl2\wedge e^x\wedge e^y,\nonumber\\
     d\tilde{G}^{2(3)}&=& 
    H\fl2\wedge e^y\wedge e^z,\nonumber\\
     d\tilde{G}^{3(3)}&=& 
    H\fl2\wedge e^z\wedge e^x,\nonumber\\
    dH\fl2&=&0.
\end{eqnarray}
which coincides with~\eqref{bB}. 
The situation corresponds to the lattice model that we have investigated in App.~\ref{sec4}. Namely, 
in a spin system with $0$-form and $1$-form global symmetries with the LSM anomaly involving these symmetries and translational ones in the two orthogonal directions, gauging $0$-form symmetries gives modulated symmetries whose dipole algebra is given in~\eqref{68}. The relations in the first and second line of~\eqref{68} corresponds to the dipole algebra introduced in~\eqref{iiii}.
Also, the relations in the third and fourth [fifth and sixth]~line of~\eqref{68} corresponds to the dipole algebra shown in~\eqref{iii}~[\eqref{iiiii}].\par
If we gauge $1$-form symmetry in a theory with~\eqref{IJK}, then we have a dual $1$-form symmetry with corresponding gauge field being $\tilde{H}\fl2$. The flatness condition of the gauge fields reads
\begin{eqnarray}
      d\tilde{H}\fl2=G^{1(1)}\wedge e^x\wedge e^y+G^{2(1)}\wedge e^y\wedge e^z+G^{3(1)}\wedge e^z\wedge e^x,\quad dG\fii1=0~(I=1,2,3).
\end{eqnarray}
This is nothing but~\eqref{aaa}. The consideration is also in line with our lattice model studied in App.~\ref{sec4}. To wit, 
we have $\mathbb{Z}_N$ analog of the dipole algebra, the one in~\eqref{iii}, ~\eqref{iiii}, and~\eqref{iiiii} which is given in the first, second, and third line of~\eqref{68}. 

\bibliography{main}
\bibliographystyle{utphys}
\end{document}